\documentclass[twocolumn]{emulateapj}
\usepackage{graphicx} 
\usepackage{natbib}
\usepackage{multirow}
\usepackage{threeparttable}
\usepackage{amsmath}

\shorttitle{Molecular Gas and Star Formation in NGC 6946}
\shortauthors{Pan et al.}

\begin{document}


\title{Molecular Gas and Star Formation Properties in the Central and Bar Regions of NGC 6946}


\author{Hsi-An Pan\altaffilmark{1,2}}
\affil{Department of Physics, Hokkaido University, Kita 10, Nishi 8, Kita-ku, Sapporo, Hokkaido 060-0810, Japan}
\author{Nario Kuno\altaffilmark{1,2,3}}
\affil{Division of Physics, Faculty of Pure and Applied Sciences, University of Tsukuba, Tennodai, Tsukuba, Ibaraki  305-8571, Japan}
\author{Jin Koda}
\affil{Department of Physics and Astronomy, Stony Brook University, Stony Brook, NY 11794-3800, USA}
\author{Akihiko Hirota\altaffilmark{2,4}}
\affil{Joint ALMA Observatory, Alonso de Cordova 3107, Vitacura, Santiago, Chile}
\author{Kazuo Sorai}
\affil{Department of Physics, Hokkaido University, Kita 10, Nishi 8, Kita-ku, Sapporo, Hokkaido 060-0810, Japan}
\and
\author{Hiroyuki Kaneko}
\affil{Nobeyama Radio Observatory, NAOJ, Minamimaki, Minamisaku, Nagano 384-1305, Japan}

\email{hapan@astro1.sci.hokudai.ac.jp}


\altaffiltext{1}{The Graduate University for Advanced Studies, 2-21-1 Osawa, Mitaka, Tokyo 181-8588, Japans}
\altaffiltext{2}{Nobeyama Radio Observatory, NAOJ, Minamimaki, Minamisaku, Nagano 384-1305, Japan}
\altaffiltext{3}{Center for Integrated Research in Fundamental Science and Engineering, University of Tsukuba, Tsukuba, Ibaraki 305-8571, Japan}
\altaffiltext{4}{National Astronomical Observatory of Japan, 2-21-1 Osawa, Mitaka, Tokyo 181-8588, Japan}

\begin{abstract} 
In this work, we investigate the molecular gas and star formation properties in the barred spiral galaxy NGC 6946 using multiple molecular lines and star formation tracers. High-resolution image (100 pc) of $^{13}$CO (1-0) is created by single dish NRO45 and interferometer CARMA for the inner 2 kpc disk, which includes the central region (nuclear ring and bar) and the offset ridges of the primary bar. Single dish HCN (1-0) observations were also made to constrain the amount of dense gas. Physical properties of molecular gas are inferred by (1) the Large Velocity Gradient (LVG) calculations using our observations and archival $^{12}$CO (1-0), $^{12}$CO(2-1) data, (2) dense gas fraction suggested by HCN to $^{12}$CO (1-0) luminosity ratio, and (3) infrared color. The results show that the molecular gas in the central region is warmer and denser than that of the offset ridges. Dense gas fraction of the central region is similar with that of LIRGs/ULIRGs, while the offset ridges are close to the global average of normal galaxies. The coolest and least dense region is found in a spiral-like structure, which was misunderstood to be part of the southern primary bar in previous low-resolution observations.    Star formation efficiency (SFE) changes by $\sim$ 5 times in the inner disk. The variation of SFE agrees with the prediction in terms of star formation regulated by galactic bar.  We find a consistency between star-forming region and  the temperature inferred by the infrared color, suggesting that the distribution of sub-kpc scale temperature is driven by star formation.

\end{abstract}



\section{Introduction}

Star formation process is intimately related to the physical properties of  molecular gas.
Physical  conditions of molecular gas  determine whether stars can form.
For example, observations of Galactic molecular gas show that star formation are often associated with dense gas \citep{Lad92}. This is true even for the galactic-scale observations as reported by \cite{Gao04}.
After the stars  form,  they heat and recycle materials back into the surrounding molecular gas,  reform  the gas by which star formation cycle can start again \citep[e.g.,][]{Oey95,Deh05,Sch12}.

In addition to the local gas conditions,  extragalactic  observations  have shown increasing signs that  molecular gas and  star formation are aware of their  galactic-scale environments. 
Dynamical properties of galaxies (e.g., bar and spiral arms) are responsible for redistributing molecular gas, controlling their formation, evolution  and ability of star formation  \citep{Ler08,Kod09,Mom10,Wat11,Hug13,Hua15}. This is  in contrast to previous studies, which generally find that star formation process  are remarkably similar across  galactic regions and galaxies \citep[e.g.,][]{Bli07,Bol08,Lad12,Don13}.

To date, studies of extragalactic molecular gas have mostly used single molecular line of $^{12}$CO because the  excitation conditions of this strong line are easily met.
However, the  low-density tracer alone is not sufficient to estimate  properties that are more intimately related to star formation, e.g., temperature and dense gas fraction.
Multi-molecular lines diagnosis is therefore indespensible  to explore the relation of molecular gas, star formation, and  galactic structures to  a greater extent.

In this work, we investigate the  physical properties of molecular gas and  star formation activity in  NGC 6946. 
These are done by analysing the newly observed isotopic molecule $^{13}$CO (1--0) (100 pc resolution) and dense gas tracer HCN (1--0), along with other archival molecular data in $^{12}$CO (1--0), (2--1) and star formation tracers in optical and infrared wavelengths.
This is  the first time that this galaxy has been observed in $^{13}$CO  with high resolution,  and one of the very few galaxies that we can perform  isotopic line mapping  down to this scale.

\begin{figure}
  \centering
    \includegraphics[width=0.45\textwidth]{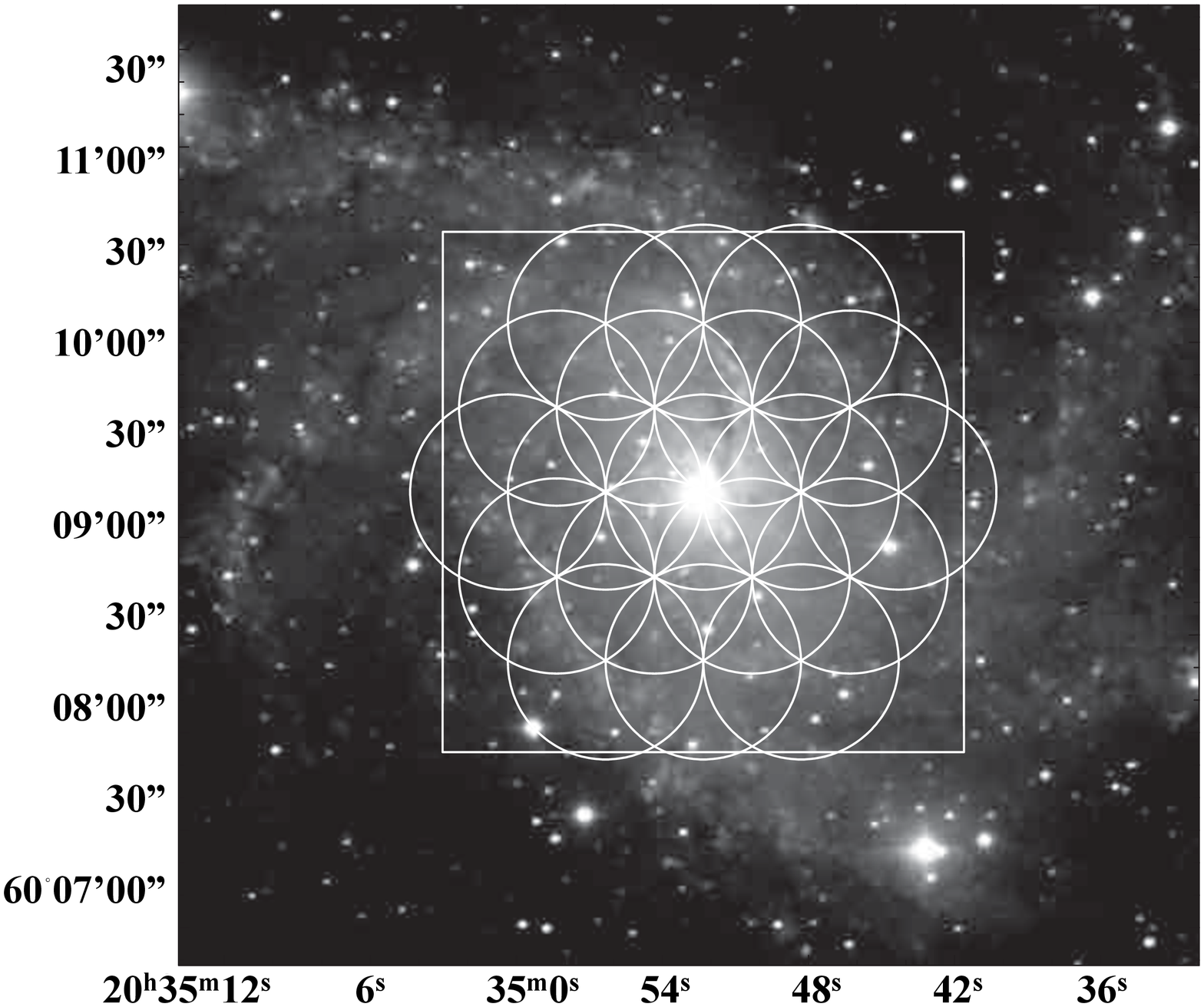}
      \caption{Observing  area overlaid on the optical $i$-band image of NGC 6946. The box indicates the observing area (160$\arcsec$ $\times$ 160$\arcsec$ or 4.3 $\times$ 4.3 kpc) of Nobeyama 45-m telescope. The center of the box corresponds to the galactic center. The 19 circles indicate the mosaic pattern of CARMA observations. The circles have a diameter of $\sim$1$\arcmin$ (1.6 kpc), corresponding to the primary beam of CARMA. The central pointing is centered at the galactic center.}
      \label{FIG_N6946opticalSmall}
\end{figure}

NGC 6946 is chosen for this work for a number of reasons. The galaxy is close by at 5.5 Mpc \citep{Tul88}, allowing us to observe in high resolution.
The face-on galaxy provides excellent viewing perspectives on the galactic structures  (Figure \ref{FIG_N6946opticalSmall}). The adopted position angle (P.A.) and inclination are 243$^{\circ}$ and 33$^{\circ}$, respectively \citep{Wal08}.
The galactic disk is characterized by four  flocculent  spiral arms,  three bars and a circumnuclear ring \citep{Sch06,Fat07}. 
The outermost  oval has  radius of  $\sim$ 7.3 kpc.
The dim dust lane (or ``offset  ridge'' towards the downstream of the galactic rotation) of the northern primary bar ($\sim$ 1 kpc) is seen in Figure \ref{FIG_N6946opticalSmall}, while the southern dust lane is not clear.
Inner region of the  primary bar  is connected to the  nuclear bar with length of $\sim$ 400 pc. The nuclear bar wraps around the starburst nucleus, forming a  circumnuclear  ring with  diameter of 20 pc. 
The disk instability and the formation of these structures have been studied through  Toomre-Q parameter \citep{Fer98,Mei04,Ler08,Rom13,Rom15}.
Data in optical and infrared  available makes the galaxy an prime target  to retrieve insight in the gas and  star formation properties.

This paper is organized as follows. 
The new observations  (molecular lines) and archival data  (molecular lines and star formation tracers) are introduced in Section \ref{sec_obs_data}.
Results of the new observations and line ratios are presented in Section \ref{sec_results}.
Galactic regions of interest are defined in Section \ref{sec_name_features}.
Section \ref{sec_physical} presents the derivation of physical properties of molecular gas.
Radial star formation efficiency is discussed in Section \ref{sec_sfr}.
Finally,  the main points of this work is summarized in Section \ref{sec_summary}.

\section{Data}
\label{sec_obs_data}

We analyze multiple molecular line transitions for investigations of physical properties of molecular gas
and optical and infrared emissions for tracing star formation activities.
We discuss our observations of $^{13}$CO (1--0) and HCN (1--0) line emission in Section
\ref{sec_obs_13co} and \ref{sec_obs_hcn}, respectively.
For $^{13}$CO (1--0), we combine single-dish observations (Section \ref{sec_obs_single_co})
and interferometric observations (Section  \ref{sec_obs_ineter}). Their combination scheme is
discussed briefly in Section \ref{sec_obs_comb}. 
The HCN (1--0) data are from single-dish observations alone. 

Archival multi-wavelength data are also used in our analyses and are presented in Section \ref{sec_archival_data}.
Table \ref{TAB_obs1213} presents a summary of the data used in this study.
In Section \ref{sec_tracers}, we will discuss physical properties that each of these emissions trace
to guide readers.

\begin{table*}
\centering
    \caption{List of newly observed and archival data sets that we have used in this work.}
    \label{TAB_obs1213}
    \begin{threeparttable}[b]
    \begin{tabular}{l*{6}{c}r}
    \hline
  Transitions/Wavelengths & Telescopes (Year/Project)&Observing modes & Resolutions &  References \\
     \hline
$^{12}$CO (1--0)        & NRO45 (2008--2010)    & mapping\tnote{$\ast$}        & 20$\arcsec$ & \citet{Don12}        \\
$^{12}$CO (1--0)        & CARMA (CANON) & mapping           & 3.2$\arcsec$\tnote{$\dag$} &\citet{Don12}          \\
 $^{12}$CO (2--1)        & IRAM30 (HERACLES) & mapping\tnote{$\ast$}           &  13.6$\arcsec$ &\citet{Ler09}          \\
 $^{13}$CO (1--0)        & NRO45 (2013)   & mapping\tnote{$\ast$}   &  20$\arcsec$  & This work  \\
$^{13}$CO (1--0)        & CARMA (CANON)    & mapping         & 3.8$\arcsec$\tnote{$\dag$} & This work  \\
HCN (1--0)        & NRO45 (2013)       & single point\tnote{$\ddag$}      & 19$\arcsec$ & This work  \\
   24 $\mu$m    & Spitzer (SINGS)      & mapping     &  5.7$\arcsec$ &\cite{Ken03}   \\
     70 and 160 $\mu$m    & Herschel  (KINGFISH)      & mapping  & 5$\arcsec$ and 12$\arcsec$  & \cite{Ken11}  \\
     H$\alpha$    & KPNO  (SINGS)       & mapping     & 3$\arcsec$ & \cite{Ken03}  \\

     \hline
    \end{tabular}
\begin{tablenotes}
\item[$\ast$] OTF observations.
\item[$\dag$] After combing with single dish data.
\item[$\ddag$] Position-switch observations.
\end{tablenotes}
\end{threeparttable}
\end{table*}

\subsection{$^{13}$CO (1--0) Observations and Data Reduction}
\label{sec_obs_13co}

Among the three data of $^{13}$CO from the NRO45, CARMA observations, we refer the combined data cube as the cube (or data or map) in this study.
The other two will be referenced explicitly as CARMA data and NRO45 data.

\subsubsection{Single Dish Observations}
\label{sec_obs_single_co}
The single dish observations of $^{13}$CO (1--0) (hereafter $^{13}$CO) were made with
the Nobeyama Radio Observatory 45m telescope (NRO45)\footnote{Nobeyama Radio Observatory is a branch of the National Astronomical Observatory of Japan, National Institutes of Natural Sciences.} in January -- February 2013.
The observations cover a 160$\arcsec$ $\times$ 160$\arcsec$ area with P.A. $=$ 0$^{\circ}$,
centering at the galactic center (see the box in Figure \ref{FIG_N6946opticalSmall}).
The observed area includes important galactic structures, such as the galactic center, galactic bar,
and the inner parts of spiral arms.
The effective beam size of NRO45 is 20$\arcsec$ at 110.2 GHz for the On-the-Fly
mapping mode \citep{Saw08},
which ensures an accurate relative flux calibration over the map.

The dual-polarization receiver TZ \citep{Asa13,Nak13} was connected to the
digital spectrometer SAM45  (Spectral Analysis Machine for the 45m telescope).
We observed with a frequency resolution of 488.28 kHz (1.3 km s$^{-1}$ at 110 GHz)
and an effective bandwidth of 1600 MHz (4356 km s$^{-1}$).
Typical system noise temperature ($T_{\mathrm{sys}}$) was 160 -- 180 K.

Each OTF map contains 33 scans in x- or y-directions and took the total of about 31 minutes.
Each scan was 20-second long with an interval of adjacent scans of 5$\arcsec$.
An OFF point 8$\arcmin$ away from the map center was observed every two scans for
the standard ON-OFF calibration. Each ON-OFF cycle (ON-ON-OFF) took 1.5 minute.
Before the observation of each map, we corrected telescope pointing by observing a point source, the SiO maser  T-Cep.
The pointing observations were performed at 43 GHz with receiver S40.
In addition, the Galactic object S140X was observed at the frequency of $^{13}$CO (1--0)
for intensity calibration once per day.
By adding up all OTF maps, the total observing time is $\sim$ 60 hours, including the overheads.

Data reduction was carried out with the package NOSTAR, which was developed for OTF data of NRO45 \citep{Saw08}.
We subtracted the spectral baseline with a one-order polynomial fit and flagged bad scans identified by eye.
OTF maps with the same scan direction (either x- or y-scan) are combined with a grid size of 6$\arcsec$,
creating two data cubes in the FITS format (x and y maps).
Finally,  PLAIT algorithm was applied to the two cubes to combine them to produce a final NRO45 cube.
This algorithm reduces scanning effect significantly.
The final NRO45 cube has the pixel scales of 6$\arcsec$ and 2.6 km s$^{-1}$ with the RMS noise of 12.1 mK in $T_{\mathrm{A}}^{\ast}$.
We adopt the main beam efficiency of 40\% for the conversion from the antenna temperature $T_{\mathrm{A}}^{\ast}$
to the main beam temperatures $T_{\mathrm{mb}}$, i.e., $T_{\mathrm{mb}}=T_{\mathrm{A}}^{\ast}/0.4$.

\subsubsection{Interferometric Observations}
\label{sec_obs_ineter}

Observations with the Combined Array for Research in Millimeter-wave Astronomy (CARMA) were made in  February -- May 2009
as a part of the CARMA-Nobeyama Nearby-galaxies (CANON) CO(1-0) survey (Koda et al. in prep.).
Some results from the $^{12}$CO (1--0) emission were published in \citet{Don12}. 
The CANON observations included the $^{13}$CO (1--0) line emission in the lower-side band of receiver,
and hence $^{13}$CO (1--0) and $^{12}$CO (1--0) were observed simultaneously.

CARMA consists of six 10 m and nine 6 m antennas.
We employed the nineteen-point hexagonal mosaic displayed in Figure \ref{FIG_N6946opticalSmall},
which covers the central part of NGC 6946.
The resultant size of the map is about 160$\arcsec$ in diameter ($\sim$ 4.3 kpc),
with the sensitivity uniform up to about 120$\arcsec$ in diameter (the central seven pointings)
and then declining to 1/2 at the 160$\arcsec$ in diameter.

Three narrow bands were used in the observations of $^{13}$CO (1--0), resulting in the total bandwidth of $\sim$108 MHz. 
The velocity channel width is 2.6 km s$^{-1}$. 
The total on-source integration time was about 21 hours including calibrators \citep{Don12}.
Bandpass, gain, and flux calibrators are 1715+096, 2015+372, and MWC349, respectively.

The CLEAN procedure is employed for deconvolution using the MIRIAD package \citep{Sau95}.
$^{13}$CO emission is often faint, and applying a spatial mask at prospective emission regions usually helps the deconvolution process.
Since $^{12}$CO is much stronger than $^{13}$CO, we expect $^{12}$CO emission always associated with $^{13}$CO emission.
We therefore made a map of  $^{12}$CO first, made a mask in channel maps, and used the mask in CLEANing the $^{13}$CO map.
The final cubes of $^{13}$CO has a velocity width of 10 km s$^{-1}$ and the RMS noise of 11 mJy beam$^{-1}$.
The beam size is 3$\farcs$29 $\times$ 3$\farcs$08 (89 pc $\times$ 83 pc) with P.A = -71.81$^{\circ}$.
  
\begin{figure*}
    \begin{minipage}{0.45\textwidth}
        \centering
		\includegraphics[width=0.9\textwidth]{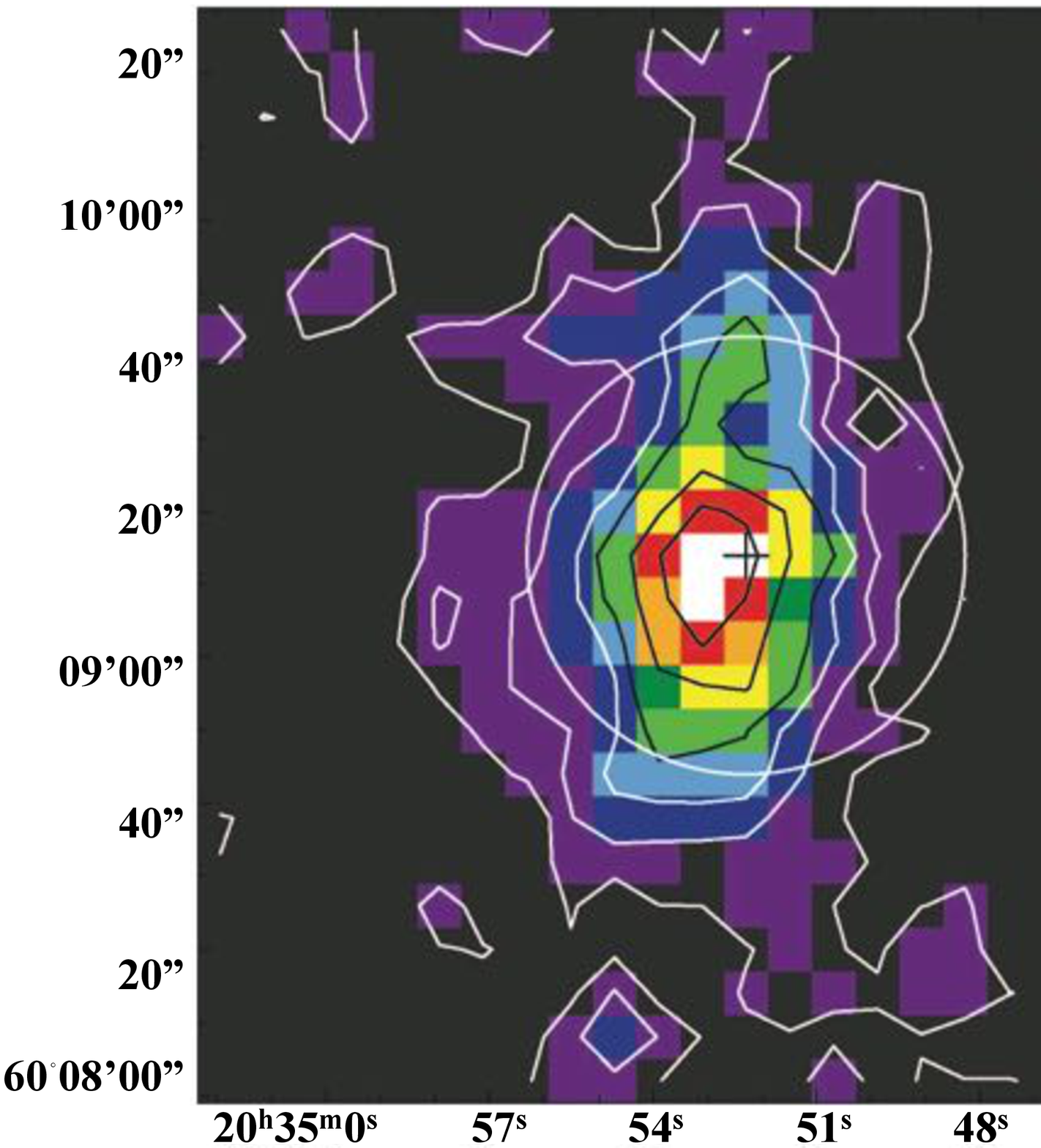}
\begin{center}
\end{center}
    \end{minipage}
    \begin{minipage}{0.45\textwidth}
    \hspace{-95pt}
        \centering
		\includegraphics[width=0.9\textwidth]{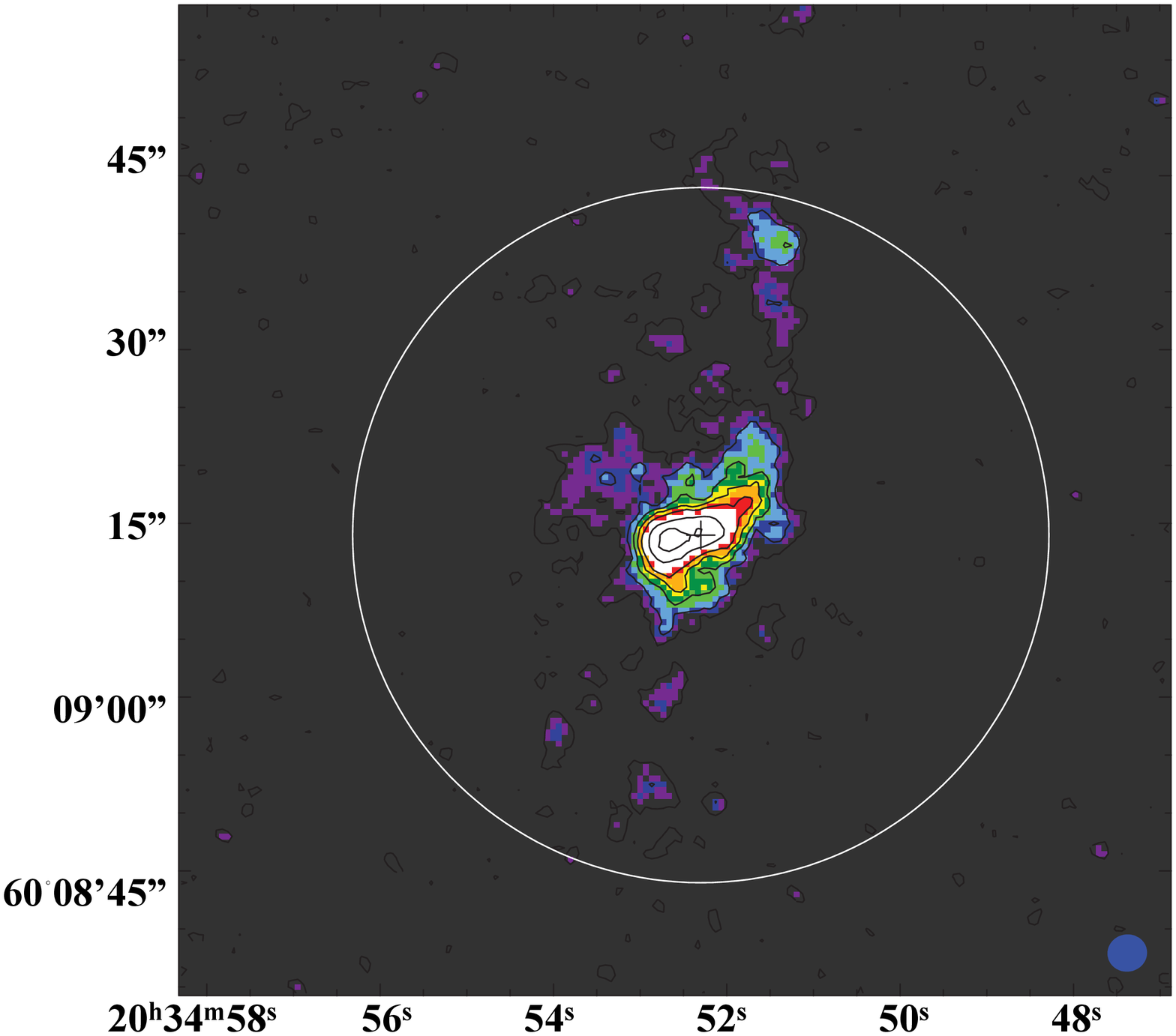}	
\begin{center}
\end{center}
    \end{minipage}
    \begin{minipage}{0.45\textwidth}
        \centering
		\includegraphics[width=0.9\textwidth]{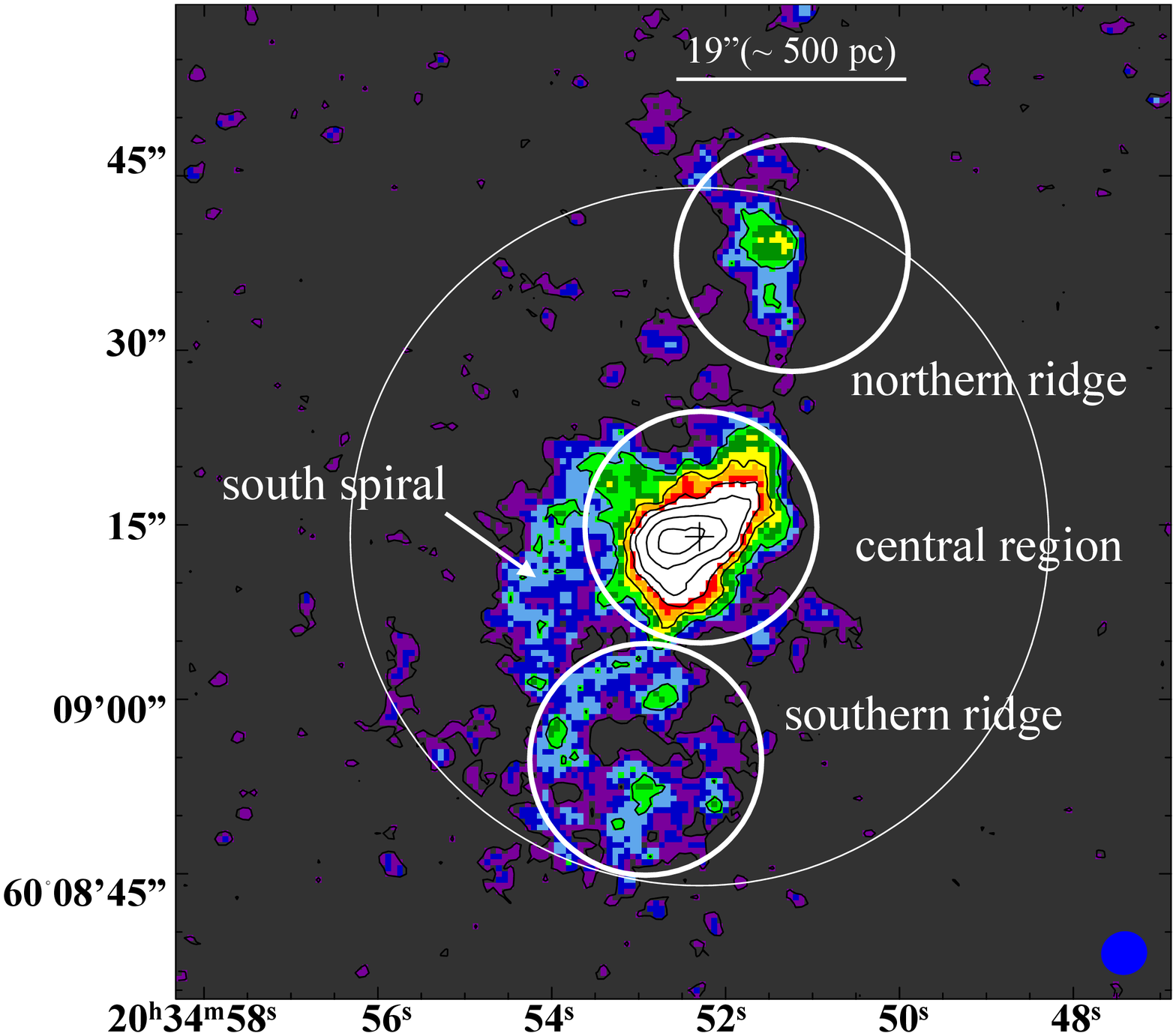}	
\begin{center}
\end{center}
    \end{minipage}
    \begin{minipage}{0.45\textwidth}
        \centering
		\includegraphics[width=0.9\textwidth]{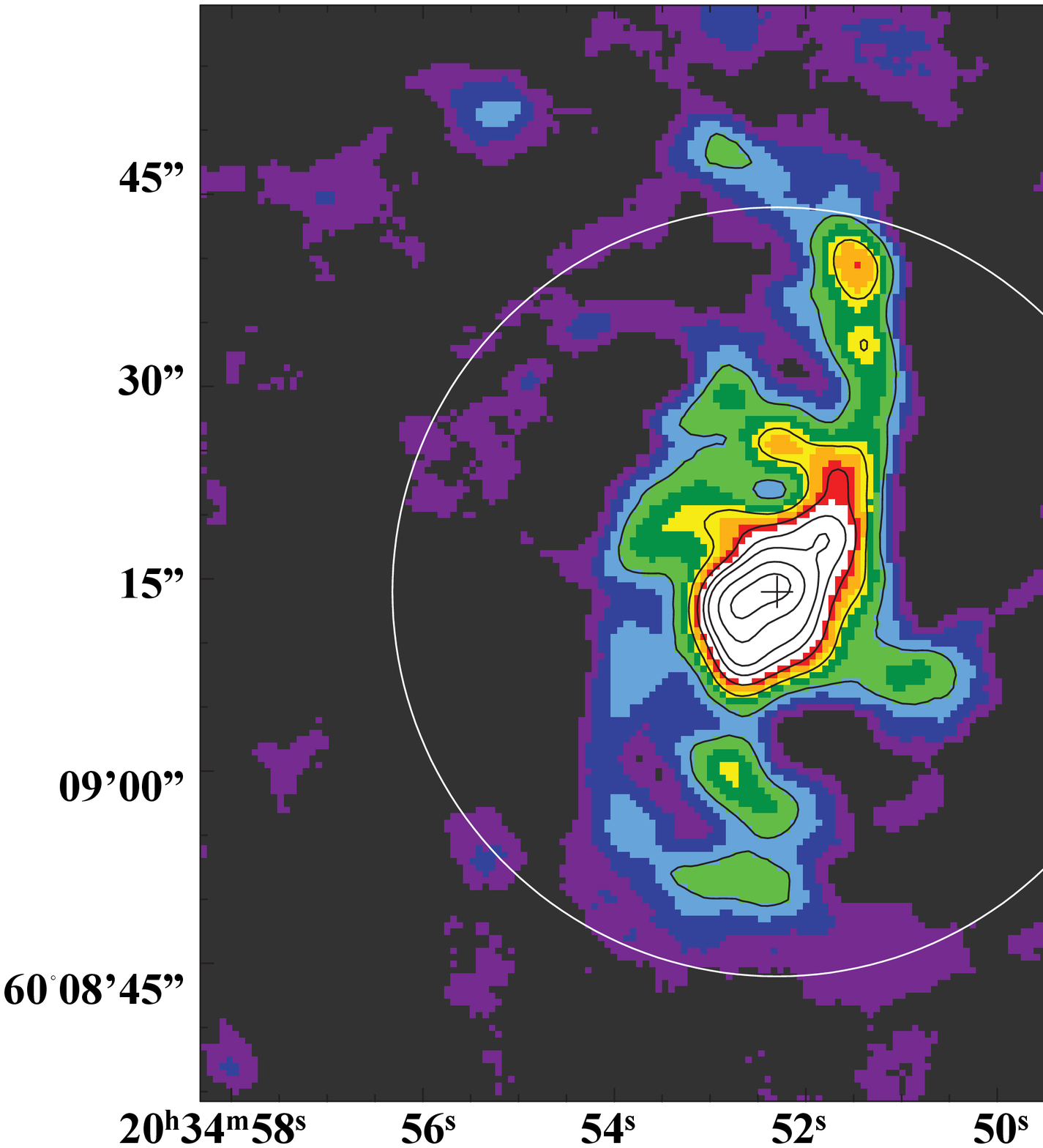}	
\begin{center}
\end{center}
    \end{minipage}
    \caption{Upper left:  Integrated intensity map of $^{13}$CO (1--0) obtained from  NRO45 in  color scale and contours. Levels of contours are 13\%, 20\%, 30\%, 40\%, 60\%, 80\% of the maximum  brightness of  22 K km s$^{-1}$.   Cross marks the galactic center at $20^{\mathrm{h}}34^{\mathrm{m}}52\fs3$, +60$\degr$8$\arcmin$14$\arcsec$. Central circle indicates 1$\arcmin$  in diameter. The beamsize of 20$\arcsec$ (540 pc) is shown in the lower-right corner. Note that this panel has different spatial scale from others in this figure.  Upper right:  Integrated intensity map of $^{13}$CO (1--0) made with CARMA data alone in color scale and contours. The contours are in steps of 10\%, 20\%, 30\%, 40\%, 50\%, 70\%, and 90\% of the maximum flux of 9 Jy beam$^{-1}$ km s$^{-1}$. The galactic center is marked with a cross. The beamsize of 3$\farcs$29 $\times$ 3$\farcs$08 (89 $\times$ 83 pc) and P.A = -71.81$^{\circ}$ is overlaid at the lower-right corner. Primary beam of CARMA is indicated with a circle, which has diameter of $\sim$ 1$\arcmin$.  Lower left:  $^{13}$CO integrated intensity map made with CARMA+NRO45  (color scale and contours). The contours are 10\%, 20\%, 30\%, 40\%, 50\%, 70\%, and 90\% of the maximum flux of 11 Jy beam$^{-1}$ km s$^{-1}$. Cross corresponds to the galactic center. The beamsize is 3$\farcs$84 $\times$ 3$\farcs$61 (103 $\times$ 97 pc) and P.A = -71.85$^{\circ}$, showing at the lower-right corner.   Three small circles indicate positions and beamsize of HCN (1--0) single dish observations. The defined regions of interest are displayed in this panel (see the text of Section \ref{sec_name_features}). Lower right: $^{12}$CO (1--0) integrated intensity map of CARMA+NRO45   (color scale and contours). Levels of contours are 10\%, 20\%, 30\%, 40\%, 50\%, 70\%, and 90\% of the maximum flux of 155 Jy beam$^{-1}$ km s$^{-1}$. The beamsize of  3$\farcs$26 $\times$ 3$\farcs$07 (88 $\times$ 83 pc) with P.A = -79.28$^{\circ}$ is superposed at the lower-right corner. Symbols are the same as in upper-right panel.}
   \label{FIG_individual_moment0s}
\end{figure*}

We first compare the maps from NRO45-alone and CARMA-alone.
The integrated intensity maps of $^{13}$CO created from NRO45-alone and CARMA-alone data are presented in the upper-left, upper-right panel of Figure \ref{FIG_individual_moment0s}, respectively.
Note that the NRO45 map covers a larger area to show a larger extent of $^{13}$CO emission.
Overall, NRO45 and CARMA capture similar structures with some differences due to the different spatial resolution
 ($20\arcsec$ vs $3\arcsec$) and the sensitivity to extended components. 
The $^{13}$CO emission appears to extend over about 60$\arcsec \times 80\arcsec$ (RA $\times$ DEC directions)
in the low resolution NRO45 map, which is resolved into more detailed structures in the CARMA-alone map.

\subsubsection{Combination Procedure of $^{13}$CO data}
\label{sec_obs_comb}
We followed \cite{Kod11} to combine the single-dish and interferometer  $^{13}$CO data.
We converted the NRO45 map into visibility data points, and then inverted the CARMA plus NRO45 visibilities
together to make dirty channel maps.
We flagged the baselines $>$4k$\lambda$ ($\sim$ 10 m) from the NRO45 visibilities because the NRO45 data
become noisier at the longer baselines and CARMA covers the long baselines sufficiently.
The dirty maps were CLEANed with MIRIAD.  
The final combined cube has the velocity resolution of 10 km s$^{-1}$ with the noise level of 14 mJy beam$^{-1}$.
The synthesized beam is 3$\farcs$84 $\times$ 3$\farcs$61 (103 pc $\times$ 97 pc) with P.A = -71.85$^{\circ}$.

We will discuss the resultant map (our default) in Section \ref{sec_results}, but for clarity of the following section, lower-left panel of Figure \ref{FIG_individual_moment0s} shows the combined $^{13}$CO map (integrated intensity map).

\subsubsection{Flux Recovered in the Combined Map}
\begin{figure}
    \begin{minipage}{0.245\textwidth}
        \centering
		\includegraphics[width=0.9\textwidth]{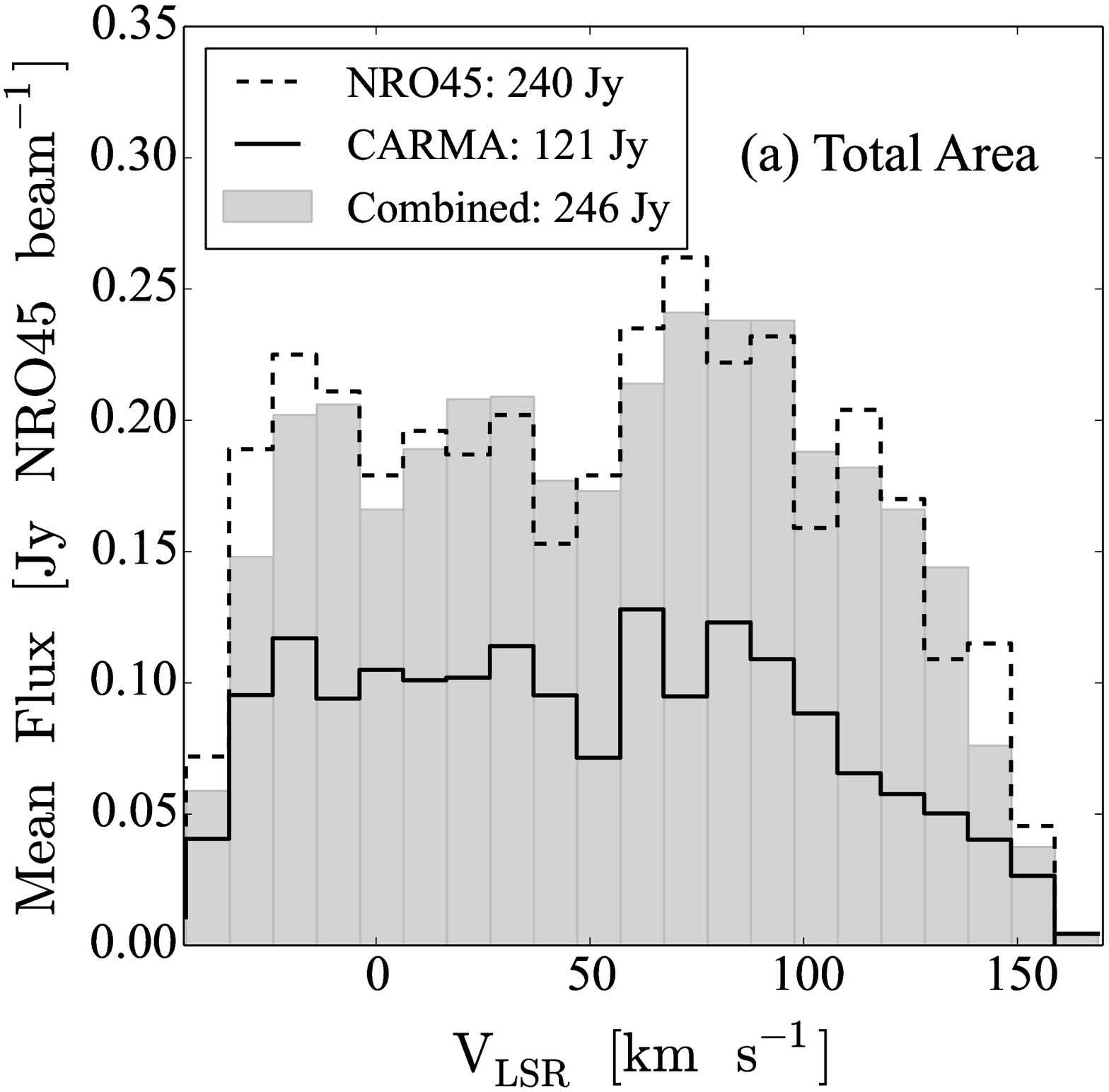}
    \end{minipage}
            \hspace{-15pt}
    \begin{minipage}{0.245\textwidth}
        \centering
		\includegraphics[width=0.9\textwidth]{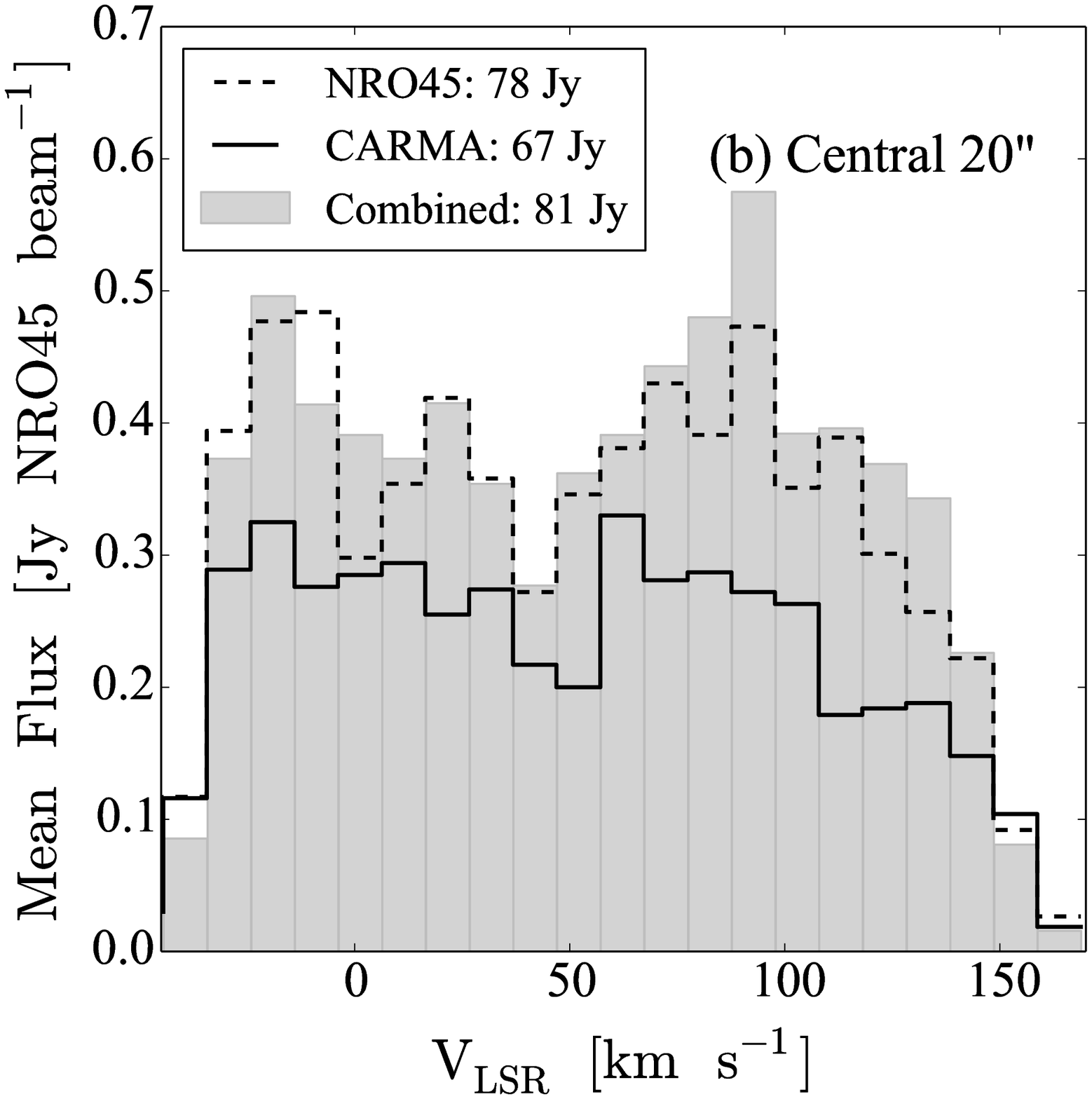}
	    \end{minipage}	
    \caption{(a) Averaged $^{13}$CO (1--0) spectra of total area (30$\arcsec$ $\times$ 70$\arcsec$ or 810 $\times$ 1900 pc).The combined spectrum is shown in solid grey histograms, and the NRO45 and CARMA spectra are shown in dashed and solid open histograms, respectively. Total flux of each measurements are presented at the upper-left corner.  (b)  Averaged spectra of the central 20$\arcsec$ ($r$ $<$ 10$\arcsec$ or 270 pc). }
   \label{FIG_fluxcheck}
\end{figure}

The fluxes are very consistent between the NRO45 and CAMRA+NRO45 cubes.
To compare the two, we smoothed the combined cube to the 20$\arcsec$ resolution, the same as that of the NRO45 cube.
Figure \ref{FIG_fluxcheck}(a) compares the average spectra over the central rectangle area (30$\arcsec$ $\times$ 70$\arcsec$)
stretched along the bar (north-south direction). Thick-line, thin-line, and shadowed histograms show the spectra of the CARMA,
NRO45, and CARMA+NRO45 cubes, respectively. The NRO45 and CARMA+NRO45 spectra are very similar overall, while
the CARMA spectrum has the flux only about 50\% of the NRO45 one.
The total integrated flux of the NRO45 and CARMA+NRO45 cubes are about 240 Jy, while the CARMA data has the total flux of $\sim$ 120 Jy,
again only 50\%; therefore, the combination of interferometer and single-dish data is very important.
Figure \ref{FIG_fluxcheck}(b) also shows a similar comparison, but within a 20$\arcsec$ aperture at the center.
The recovered flux by CARMA-alone is $\sim$ 80\% in total flux with respect to NRO45.

\subsection{HCN (1--0) Observations and Data Reduction}
\label{sec_obs_hcn}

Observations in HCN(1-0) were carried out in January 2013 using NRO45.
We observed only three selected positions in the HCN(1-0) line emission, because 
this emission is weak \citep[e.g., $I\mathrm{_{HCN}}/I\mathrm{_{^{13}CO}}<$ 0.3 in galactic disks; ][]{Mat10}.
These pointed observations reveal the amount of dense gas in the regions of interest.
The three positions are the galactic center at  ($20^{\mathrm{h}}34^{\mathrm{m}}52\fs3$, +60$\degr$9$\arcmin$14$\arcsec$), and two off-center regions at  ($20^{\mathrm{h}}34^{\mathrm{m}}51\fs3$, +60$\degr$9$\arcmin$38$\arcsec$) and  ($20^{\mathrm{h}}34^{\mathrm{m}}52\fs9$, +60$\degr$8$\arcmin$55$\arcsec$) (circles in the lower-left panel of Figure \ref{FIG_individual_moment0s}).

We used the receiver TZ and spectrometer SAM45 and employed the position-switch mode of observations.
The frequency resolution is 488.28 kHz (1.6 km s$^{-1}$ at 88.6 GHz),
and the effective bandwidth of 1600 MHz ($\sim$ 5416 km s$^{-1}$).
An OFF point 8$\arcmin$ away from the target was observed every 15 seconds for the ON-OFF calibration.
We checked telescope pointings every 45 minutes with the SiO maser T-cep.
$T_{\mathrm{sys}}$ was about 140 K during the observations.
A standard flux calibrator S140 was observed once per day. 
The total on-source integrated time is about one hour at each position.
The NRO45 beamsize at the HCN (1--0) frequency is about 19$\arcsec$, roughly comparable to
the beamsize at CO of 20$\arcsec$ after regridding (smoothing) the OTF data onto the grid of the final data cube.

Data reduction was carried out with the NEWSTAR package developed at the Nobeyama observatory.
We subtracted spectral baselines from each spectrum using a linear fit and flagged some bad spectra (with non-flat baselines).
We binned the spectra, and the final spectra have a velocity resolution of 13 km s$^{-1}$.
We detected the emission from the galactic center at an 18 $\sigma$ significance, and   5 -- 6 $\sigma$ significance at the off-center regions.

\subsection{Archival Data}
\label{sec_archival_data}

We analyze $^{12}$CO (1--0),  $^{12}$CO (2--1), $^{13}$CO (1--0),  HCN (1--0), 70$\mu$m, and 160$\mu$m  to probe physical conditions of molecular gas in NGC 6946.
The $^{13}$CO (1--0) and HCN (1--0) emission data are from our own observations as discussed above,
and $^{12}$CO (1--0) and  $^{12}$CO (2--1) are obtained from
the CARMA-Nobeyama Nearby-galaxy Survey \citep[CANON; Koda et al. in prep.;][]{Don12}
and The HERA CO-Line Extragalactic Survey \citep[HERACLES; ][]{Ler09}, respectively.

To investigate star formation activities, we use the archival data of  Spitzer 24$\micron$ and
H$\alpha$ recombination line emissions.
The 24$\mu$m and H$\alpha$ images are taken from the Spitzer Infrared Nearby Galaxies Survey \citep[SINGS; ][]{Ken03}.
We also use the 70 $\mu$m and 160 $\mu$m images from the Key Insights on Nearby Galaxies: a Far-Infrared Survey with Herschel \citep[KINGFISH;][]{Ken11} to infer the trend of temperature of dust and gas.

\subsection{Physical Properties Traced by Each Emission}
\label{sec_tracers}

This study compares multi-wavelength data to investigate the physical conditions of gas and star formation activities
in NGC 6946. It is perhaps useful to summarize what each of these multi-wavelength data traces physically.

\subsubsection{Molecular Line Emissions}
\label{sec_tracers_molecular}
The $^{12}$CO(1-0) line is often used to trace the amount of bulk molecular gas.
$^{12}$CO is the second most abundant molecule after H$_{2}$.
The temperature equivalent to the first energy level of rotational transition is $\sim$ 5.5 K, therefore the $J=1$ level is
always populated very well for the the typical temperature of molecular gas ($\sim$ 10 K).
Its typically high opacity prevents photons  escaping efficiently from the emitting area, which drops the effective
critical density for excitation low \citep[$\sim 300 \rm\, cm^{-3}$;][]{Sco74}.
The average density within molecular clouds is comparable to this density \citep{Sol87}
and therefore, the bulk of molecular gas within molecular clouds emits $^{12}$CO(1-0) emission efficiently.

The temperature equivalent to the $J=2$ energy level is $\sim 15$ K, slightly above the typical gas temperature,
and hence $^{12}$CO (2--1) is sensitive to slight enhancements in gas temperature or density of the bulk molecular gas \citep{Kod12}.
$^{12}$CO(1-0) is often used as a tracer of molecular gas mass \citep[][and references therein]{Bol13}, even though $^{12}$CO is generally optically thick.
The velocity dispersion of molecular clouds is almost always larger than the thermal line width, and in fact, this optically-think line can
trace the entire volume within the clouds, and thus their mass. We use the $^{12}$CO(1-0) emission for calculation of molecular gas mass.

The $^{13}$CO (1--0) emission (hereafter $^{13}$CO) is also used to trace bulk molecular gas.
It is typically optically-thin, compared to $^{12}$CO(1-0), and therefore, its effective critical density ($\sim 2\times 10^3 \rm\, cm^{-3}$)
is an order of magnitude greater than that of $^{12}$CO, tracing slightly denser gas.
Abundance ratio of $^{12}$CO to $^{13}$CO is about 40 -- 60, therefore $^{13}$CO emission is significantly weaker than $^{12}$CO.

The HCN (1--0) emission (hereafter HCN) is often used as a tracer of star-forming dense cores within molecular clouds.
It has a high critical density ($\sim$ 10$^{5-6}$ cm$^{-3}$), and the HCN emission, even unresolved, should be
coming selectively from the very dense regions within molecular clouds. The connection between these dense regions 
and star formation activities are seen in the linear correlation between HCN and tracers of star formation rate.
We use HCN to constrain the amount of dense gas.
Enhanced HCN emission around AGNs \citep[e.g.,][]{Ima07,Izu13} may be a source of confusion when the galactic
center is the focus of study, but the mechanism of the enhancement is irrelevant here, since
NGC 6946 has no appreciable supermassive black hole \citep{Kor07,Kor10}.

\subsubsection{H$\alpha$ and Infrared Emissions}

H$\alpha$ and infrared emissions are often used to trace the intensity of star formation.
Both types of emissions are the second product of recently-formed young stars, with
H$\alpha$ emission from the gas ionized by UV photons from young, massive stars
and infrared, such as 24$\mu$m emission tracing the thermal radiation from dusts heated predominantly by young stars  at the age up to $\sim$ 10 Myr \citep{Cal05}.

Each emission has its own advantages and disadvantages in estimating star formation rate (SFR).
H$\alpha$ typically provides a high spatial resolution, but suffers from dust extinction.
The bottleneck of current infrared data is its relatively low spatial resolution, though the extinction is not so much a problem at the infrared wavelengths.

The combination of the two may complement each other and provide a more accurate estimate of SFR \citep[e.g., ][]{Cal05,Ken07,Ca12}, although the spatial resolution may be an issue here since it needs to be adjusted to the lowest infrared resolution.
Here, we use 24 $\mu$m and H$\alpha$ emissions to gauge star formation activities.

Infrared color is used to infer dust temperature.
Young massive stars contribute to the SED at shorter wavelengths, producing a peak around $\sim$ 60 $\mu$m, while low-mass stars contribute to the SED at longer wavelengths, generating another peak around $\geq$ 160 $\mu$m \citep{Ca12}.
Therefore,  ratio of the fluxes around the two peaks provides a probe of dust and gas temperature. 
In this work, we use 70 $\mu$m to 160 $\mu$m flux ratio or color to trace the temperature variation.

\section{Observational Results}
\label{sec_results}
\subsection{$^{13}$CO  Observations}
\label{sec_co_morph}
\subsubsection{Channel Map}
\begin{figure*}
  \centering
    \includegraphics[width=0.9\textwidth]{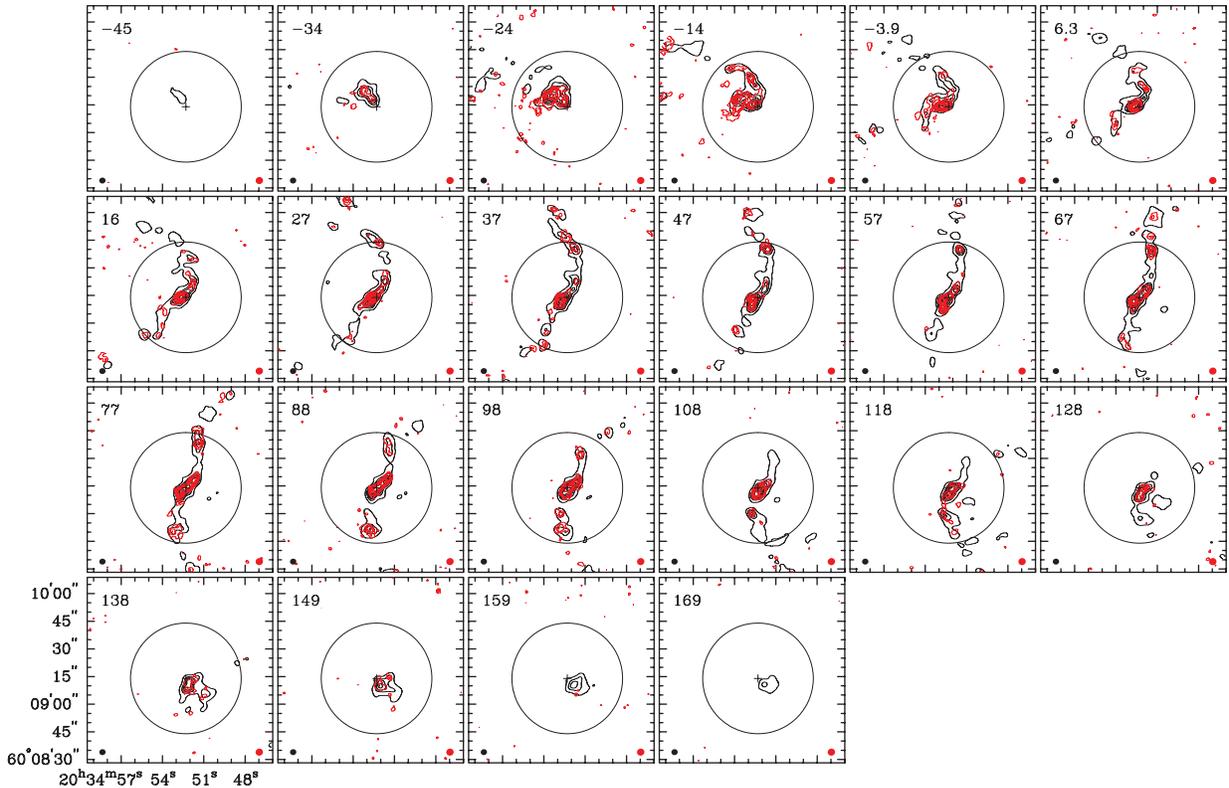}
      \caption{Channel maps of combined CO (1--0) lines. $^{13}$CO is shown in red contours and $^{12}$CO in black. Beamsize of $^{13}$CO (3$\farcs$84 $\times$ 3$\farcs$61 or 103 $\times$ 97 pc, P.A = -71.85$^{\circ}$) and $^{12}$CO (3$\farcs$26 $\times$ 3$\farcs$07 or 88 $\times$ 83 pc, P.A = -79.28$^{\circ}$) are indicated at the lower-right and lower-left corner with colors of red and black, respectively.  The galactic center is marked with a cross in each panel. Area of central 1 $\arcmin$ is highlighted with a circle. Values at the upper left corner denotes the velocity of each channel in unit of km s$^{-1}$. $^{13}$CO is plotted with a contour step of 25\%, 35\%, 45\%, 55\%, 75\%, and 90\% of the maximum flux of 0.1 Jy beam$^{-1}$. Contours of $^{12}$CO are 15\%, 25\%, 35\%, 45\%, 55\%, 75\%, and 90\%  of the maximum flux of 1.6 Jy beam$^{-1}$.}
      \label{FIG_channel}
\end{figure*}

The channel maps of $^{13}$CO are displayed in Figure \ref{FIG_channel} with red contours, while
the $^{12}$CO contours (black) are also plotted for reference (The $^{12}$CO data also include CARMA and NRO45 data).
The galactic center is marked with a cross in each channel.
There is a central peak around the galactic center.
The central peak has a velocity width of  $\sim$ 180 km s$^{-1}$, ranging from --34 km s$^{-1}$ to 149 km s$^{-1}$.
Apart from the central component, two elongated structures emerging from the galactic center are seen, extending toward the north and the south, respectively. 
Both sides show sharp velocity gradients across the elongated structures. 
Such pattern is commonly seen in galactic bars \citep[e.g.,][]{Kod06}.
Emission in the northern region emerges from $\sim$ --14 km s$^{-1}$, spreading over $\sim$ 112 km s$^{-1}$.
Emission in the southern region emerges from $\sim$ --39 km s$^{-1}$, spreading over $\sim$ 110 km s$^{-1}$.

The spatial and velocity distributions of $^{13}$CO and $^{12}$CO emission are similar in all channels.
All $^{13}$CO peaks have  counterparts in $^{12}$CO. 
This is a natural result as mentioned in \S\ref{sec_tracers}.
On the contrary, $^{13}$CO is absent at some $^{12}$CO peak (e.g., 67 km s$^{-1}$).
It is due to either the detection limit or an insufficient density for $^{13}$CO excitation.

\subsubsection{Integrated Intensity Maps}
\label{sec_mom0}
The CARMA+NRO45 map clearly shows the central component in details, resolving the central concentration elongated
toward north-west to south-east directions. The semi-major axis of the elongation is about 8$\arcsec$ ($\sim$ 220 pc),  corresponding to the secondary (nuclear) bar formed via local gravitational instability in the disk \citep{Elm98,Sch06,Fat07,Rom15}.
Structure inside  the secondary  bar is not resolved in our map.
Emission appears to extend toward north from the edge of this elongation reaching the radius of $\sim$40$\arcsec$ toward north.
We call this extension   ``Northern Ridge'', which has a bright peak near the CARMA field of view.

The $^{12}$CO map from CARMA+NRO45  is displayed in the lower-right panel of Figure \ref{FIG_individual_moment0s} for comparison.
The $^{13}$CO and $^{12}$CO maps are generally similar, though $^{12}$CO shows more continuous extension overall. 
For example, the emission extends smoothly along the northern ridge in $^{12}$CO, but shows a gap in $^{13}$CO at around
the radius of 16$\arcsec$ -- 18$\arcsec$ which is between the central concentration and the northern ridge.
The lack of $^{13}$CO emission could be due to an insufficient sensitivity for detection of this weak line, but we also point out that the level of $^{13}$CO emission must be lower than what is expected in assumption of
the $^{12}$CO/$^{13}$CO line ratio of 15, i.e., the average over the area of significant $^{13}$CO detection within the northern ridge.
Assuming this ratio, the expected $^{13}$CO flux is $\sim$ 1.8 Jy beam$^{-1}$ km s$^{-1}$  in this gap, which should be detected at the 4 $\sigma$ significance.
Therefore, the $^{12}$CO/$^{13}$CO ratio is enhanced at the connection between the central concentration and northern ridge, suggesting a change of molecular gas properties along the bar. 
Such enhancement has also been reported in the strong bar galaxy NGC 7479 by \cite{Hut00}.

The main difference between the CARMA+NRO45, NRO45-alone and CARMA-alone  maps appears at the south side of the galaxy.
The CARMA+NRO45 maps show a curved spiral-like pattern at the southeast of the center, which appears as extended emission in the NRO45 map, but appear only as distributed/not-connected emission peaks in the CARMA map.
Some of the extended emission is not detectable in the CARMA-alone map, which is recovered by the combination.

\subsection{HCN Observations}
\label{sec_hcn_results}

\begin{figure*}
    \begin{minipage}{0.32\textwidth}
        \centering
		\includegraphics[width=0.9\textwidth]{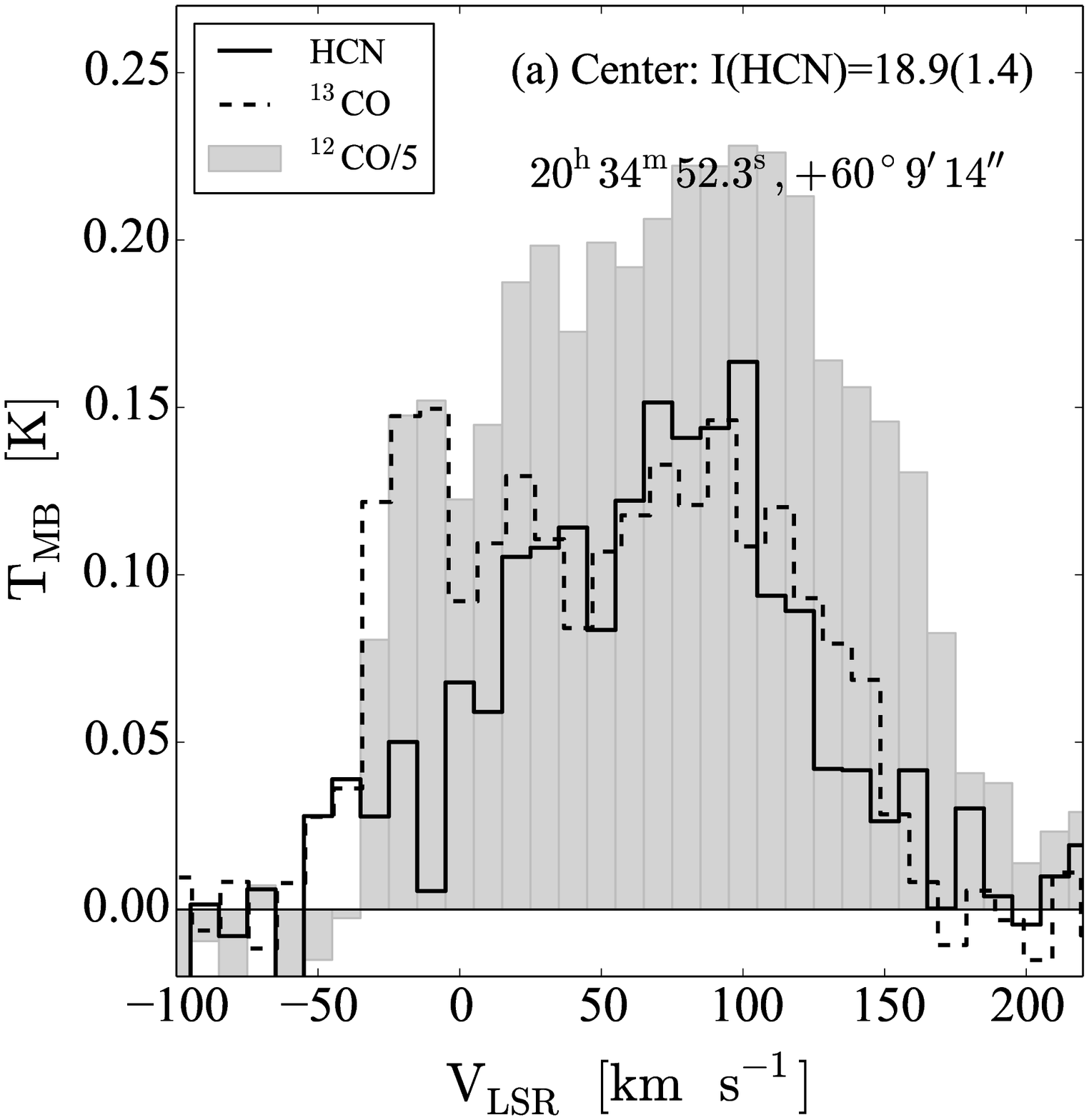}
    \end{minipage}
    \begin{minipage}{0.32\textwidth}
        \centering
		\includegraphics[width=0.9\textwidth]{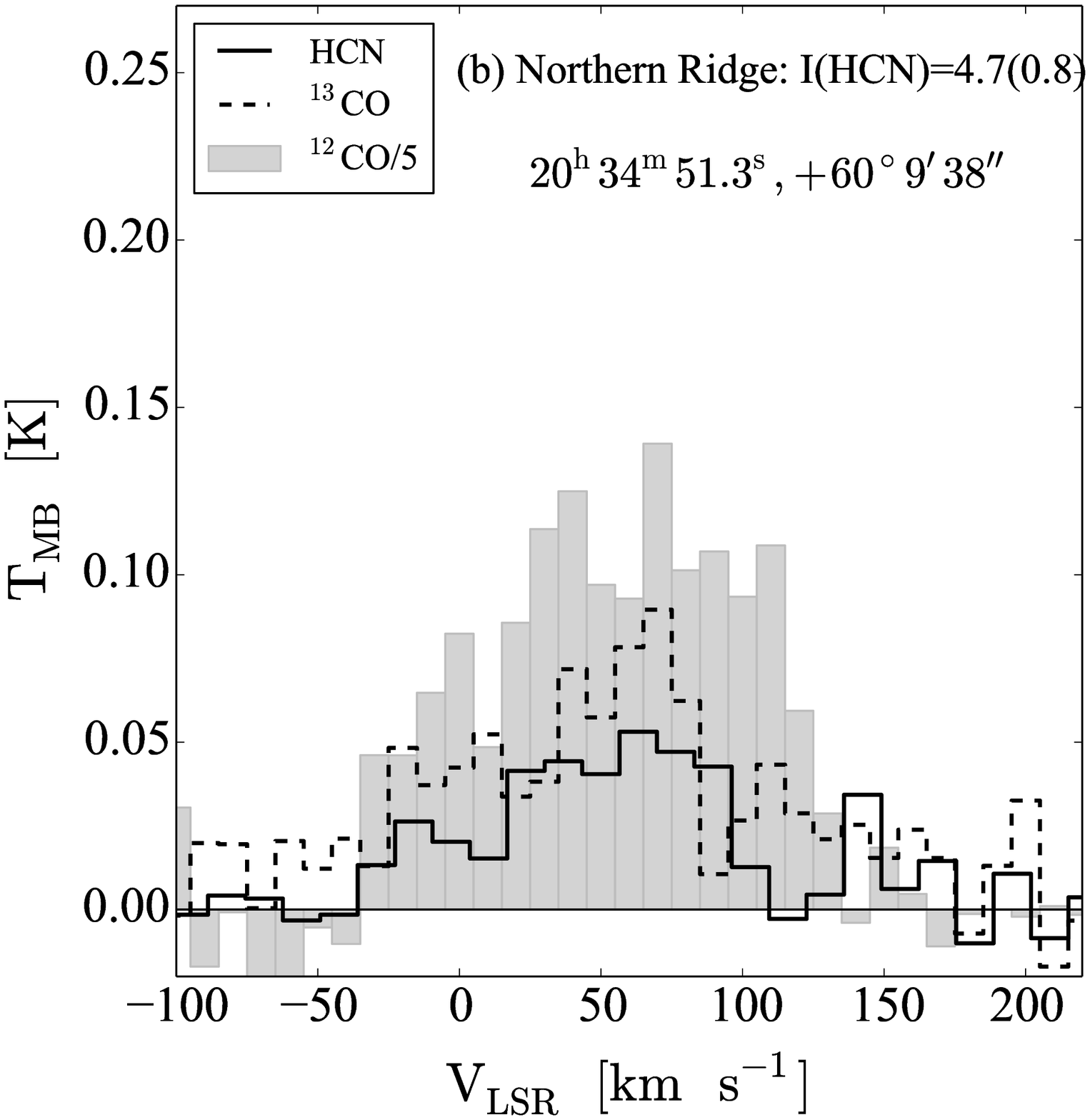}
	    \end{minipage}	
	        \begin{minipage}{0.32\textwidth}
        \centering
		\includegraphics[width=0.9\textwidth]{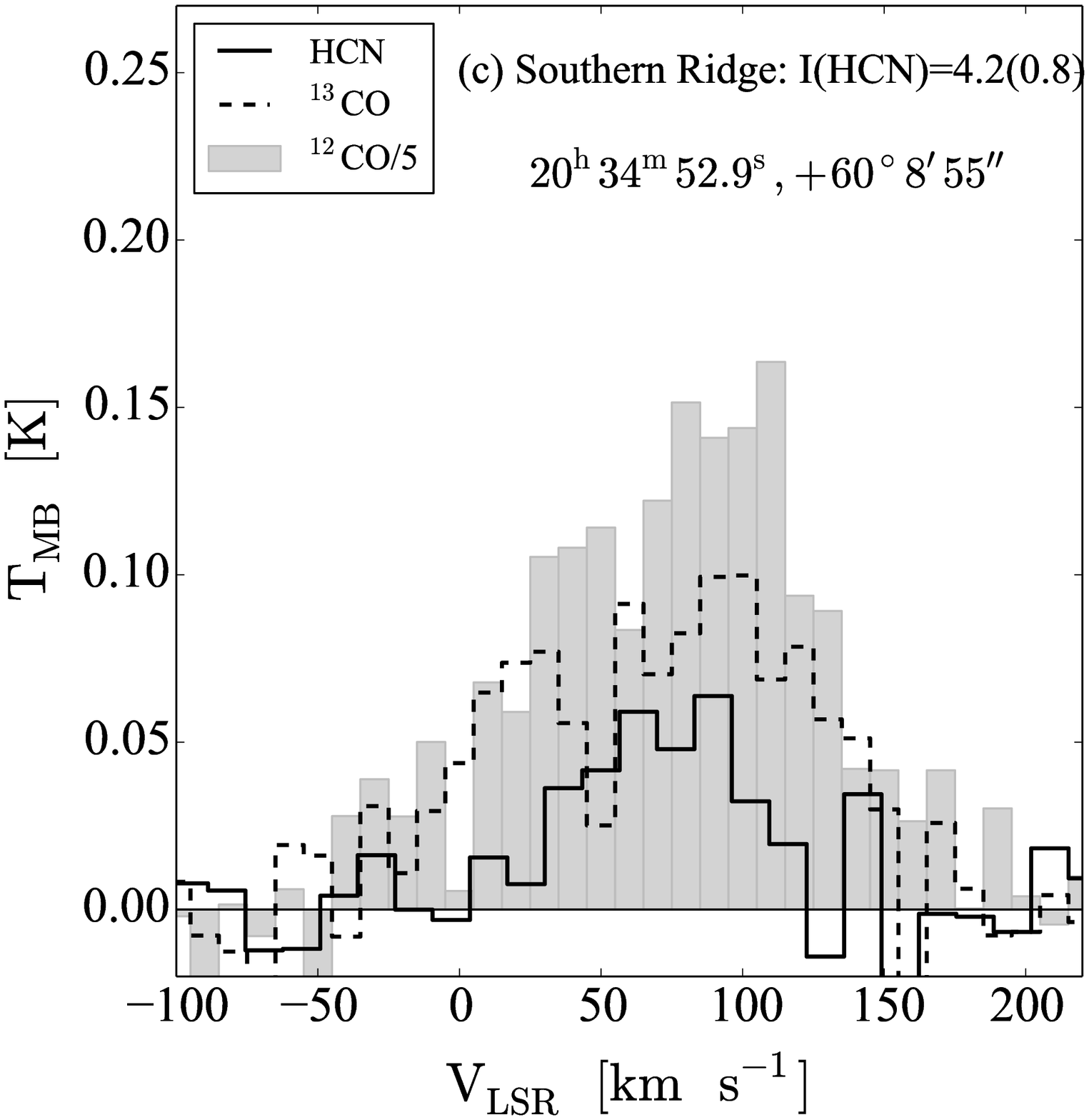}
	    \end{minipage}
    \caption{Single dish spectra of HCN (1--0) (solid open histograms), $^{13}$CO (1--0) (dashed open histograms), and $^{12}$CO (1--0) (solid grey histograms). Intensity of $^{12}$CO has been divided by a factor of five.  Velocity resolutions of HCN and CO spectrum are 13 km s$^{-1}$ and 10 km s$^{-1}$, respectively. All spectra correspond to a region of $\sim$ 19$\arcsec$ in this figure. Coordinates of the single point observations are provided in each panel. (a) Spectra of the central region, centering at the galactic center. (b) Spectra of the northern ridge. (c) Spectra of the southern ridge. }
   \label{FIG_HCN_spec}
\end{figure*}

Figure \ref{FIG_HCN_spec}(a), (b), and (c) compares HCN spectra against $^{12}$CO and $^{13}$CO spectra (NRO45 alone)
at the center, north off-center, and the south off-center regions, respectively.
Their overall shapes are similar to each other, except that the spectra at the south off-center position show a slight difference.
At this position the HCN spectrum shows a single peak, while the two CO lines show two peaks at $\sim$ 40 and 80 km s$^{-1}$,
suggesting that CO and HCN trace different gas components.

The HCN integrated intensity is 18.9 $\pm$ 1.4, 4.7 $\pm$ 0.8, and 4.2 $\pm$ 0.8 K km s$^{-1}$ at the center, north, and the south off-center regions, respectively.

Luminosity ratio of $L'_{\mathrm{HCN}}$/$L'_{\mathrm{^{12}CO}}$ of the galactic center, north , and the south off-center regions  are 0.111 $\pm$ 0.011, 0.065 $\pm$ 0.013, and 0.051 $\pm$ 0.012, respectively.
Luminosity ratio of the off-center positions are close to the  global average values of normal galaxies  while $L'_{\mathrm{HCN}}$/$L'_{\mathrm{CO}}$ of the central region lies between the mean value of LIRGs  and ULIRGs \citep{Gao04}, where the definition of  LIRG and ULIRG  are $10^{11}L_{\sun}<L_{\mathrm{IR}}\leq0.8\times10^{12}L_{\sun}$ and $L_{\mathrm{IR}}\geq 10^{11.9}L_{\sun}$, respectively.

\subsection{Line Ratio of CO lines}
\label{sec_co_ratios}
\begin{figure}
  \centering
    \includegraphics[width=0.45\textwidth]{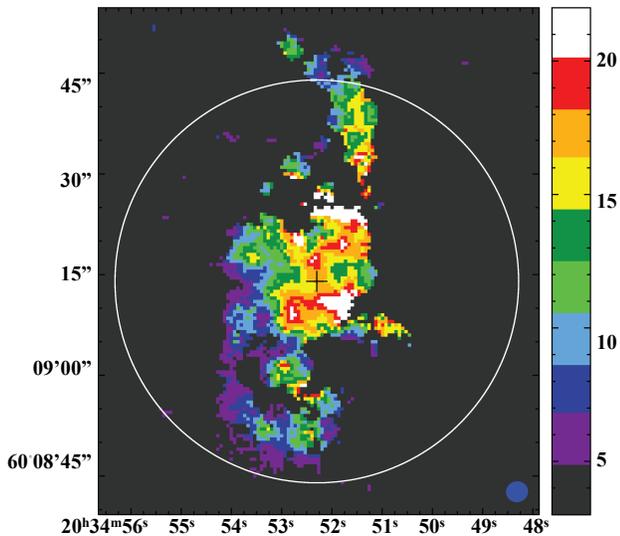}
      \caption{$R_\mathrm{10}$ ($I_{\mathrm{^{12}CO(1-0)}}/I_{\mathrm{^{13}CO(1-0)}}$) ratio map of central 1$\arcmin$. The galactic center is marked with a cross. Central 1$\arcmin$ is indicated by a circle.
Beamsize  is plotted at the low-right corner.}
      \label{FIG_ratiomap}
\end{figure}

Emission line ratio often provides an idea of physical properties of molecular gas.
The $^{12}$CO to $^{13}$CO intensity ratio ($R_\mathrm{10}$) is presented in Figure \ref{FIG_ratiomap}.
This ratio varies by a factor of three -- the maximum ($\sim$ 17) around the galactic center to the minimum ($\sim$ 6) in the spiral-like ridge,
which covers the large range observed in typical Galactic molecular clouds ($< 10$) to starburst galaxies and galaxy mergers \citep[10 -- 20, though sometimes $>$10 -- 20; ][]{Sol79,Aal95,Tan98,Pag01,Tan11}.

\begin{figure*}
    \begin{minipage}{0.45\textwidth}
        \centering
		\includegraphics[width=0.9\textwidth]{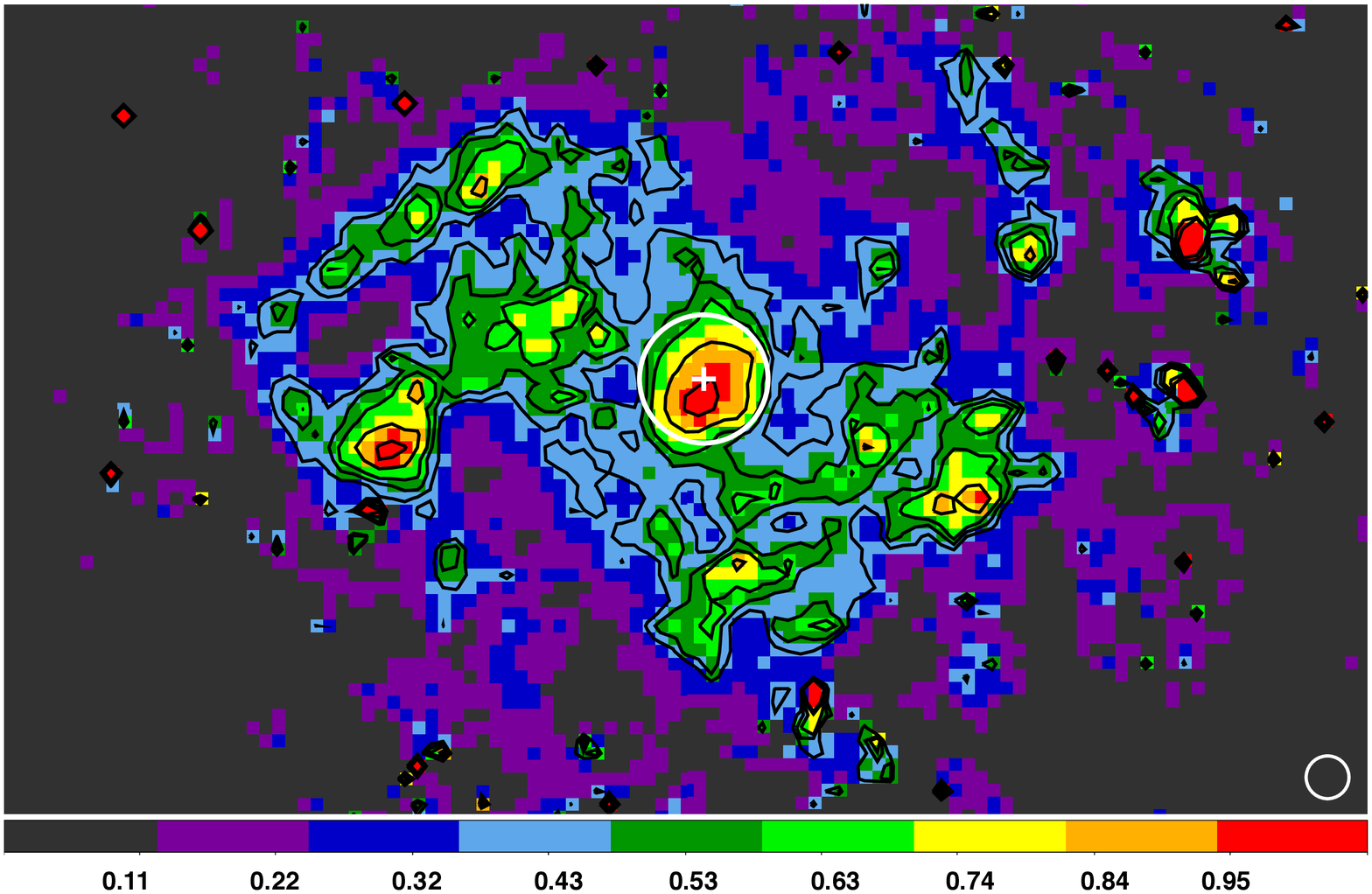}
    \end{minipage}
    \begin{minipage}{0.45\textwidth}
        \centering
		\includegraphics[width=0.9\textwidth]{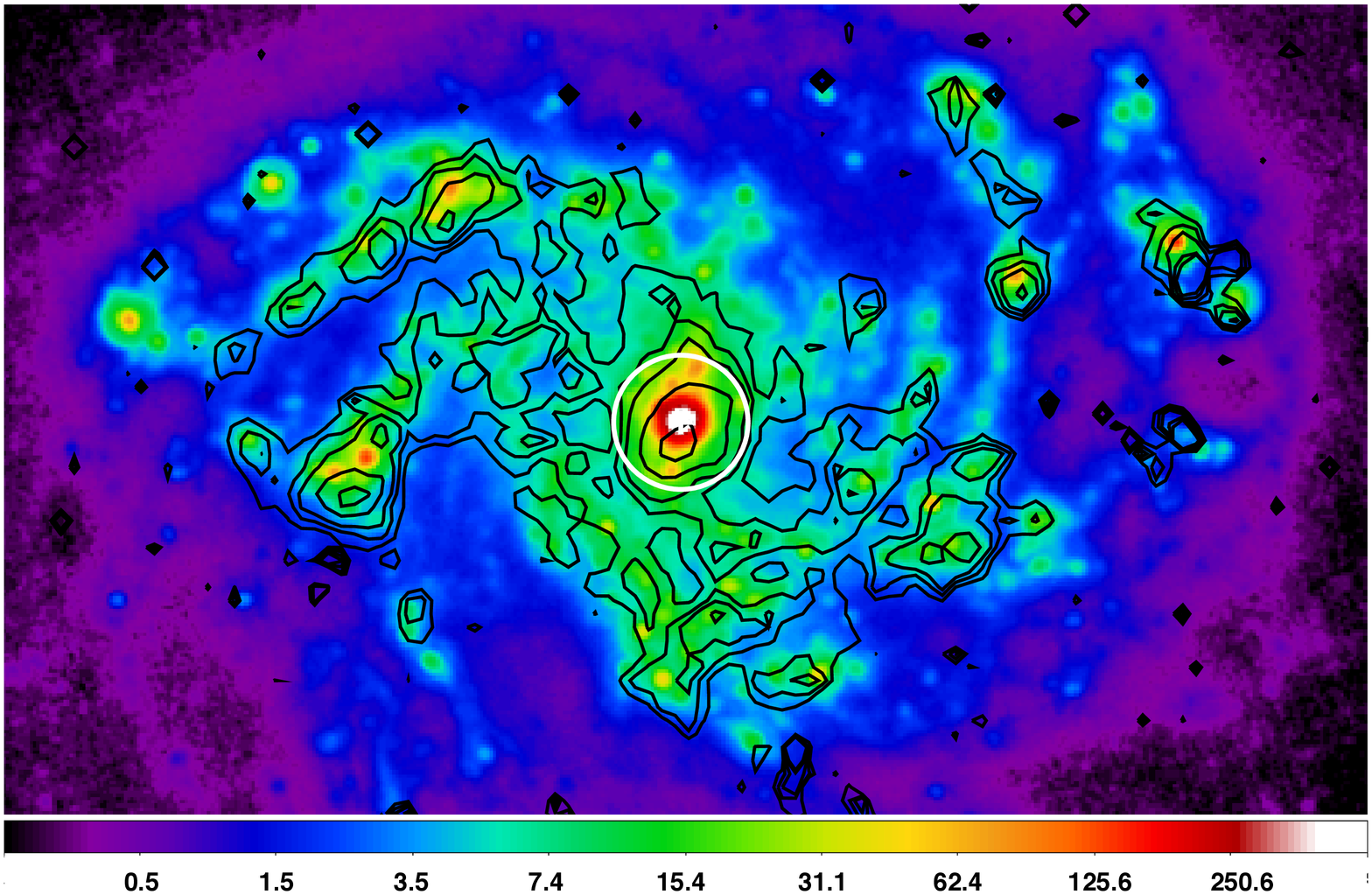}
	    \end{minipage}	
    \caption{$R_\mathrm{21}$ ($I_{\mathrm{^{12}CO(2-1)}}/I_{\mathrm{^{12}CO(1-0)}}$) map made with single dish observations of NRO45 and IRAM30. Both contours and color scale represent $R_\mathrm{21}$. Contours run in steps of 0.4, 0.5, 0.6, 0.8, and 1.0.  Central circle indicates 1$\arcmin$ in diameter. Cross marks the galactic center. Beamsize of 20$\arcsec$ is over-plotted at the lower-right corner.  Right: $R_\mathrm{21}$ map (contours) overlaid on Spitzer 24$\mu$m image (color scale). Steps of contour are the same as in the left panel. The 24$\mu$m image is shown with original resolution of $\sim$ 5$\arcsec$.  }
   \label{FIG_R21}
\end{figure*}

Variable $^{12}$CO (2--1)/(1--0) ratio ($R_\mathrm{21}$) is also observed in NGC 6946.
Left panel of Figure \ref{FIG_R21} presents the $R_\mathrm{21}$ map of entire galaxy.  
Since there is only single dish $^{12}$CO (2--1) image, we calculate $R_\mathrm{21}$ with single dish $^{12}$CO (1--0).
$R_\mathrm{21}$ map has a resolution of  20$\arcsec$, which is the resolution of NRO45 data of $^{12}$CO (1--0).
The ratio  map shows a central oval with $R_\mathrm{21}$ $\approx$ 1. 
The orientation of this oval  is consistent with the unresolved nuclear bar. 
$R_\mathrm{21}$ is about 0.5 -- 0.8 at the spiral arms, and 0.3 -- 0.5 at the inter-arm regions. 
These ratios are comparable to that in the Milky Way and nearby galaxies \citep[e.g.,][]{Sak97,Oka98,Saw01,Kod12}.

Right panel of Figure \ref{FIG_R21} compares the spatial distribution of $R_\mathrm{21}$ and  star forming regions  traced by Spitzer 24$\mu$m. 
The high $R_\mathrm{21}$ is spatially correlated with the location of stars.

\section{Gas Structures from CO Observations}
\label{sec_name_features}
The morphology of molecular gas in NGC 6946 resembles the typical  barred spiral galaxies \citep[e.g.,][]{She02,Kod06}. 
Barred spiral galaxies often show a central concentration of gas and offset ridges that extend from the central concentration along the leading side of the bar. Such structures are often reproduced in numerical simulations \citep[e.g.,][]{Ath99}, and analytical gas orbit models \citep{Wad94,Sak99,Kod02,Kod06}.
The bar of this galaxy runs in the north-south direction in an optical image. The CO map shows the central concentration with a major axis of $\sim$ 20$\arcsec$ ($\sim$ 540 pc), and two ridges with  length $\sim$ 20 -- 30$\arcsec$, running from the central concentration toward the north and south directions with the southern ridge appear fragmented.

For clarity of the rest of discussions, we define molecular gas structures based on the morphology in the CO map.
We call the central concentration (oval structure with the long extension of 20$\arcsec$ elongated toward southeast-northwest  direction) as \emph{central region}. This definition of the central region includes the secondary bar as well as the galactic center
(lower-left panel of Figure \ref{FIG_individual_moment0s}).
The two ridges extending from the central region toward north and south are called \emph{the northern ridge} and \emph{southern ridge}, respectively.
From a closer look at the distribution and kinematics  we notice that there is an additional
component superposed on the southern ridge. This component resembles a spiral arm, originating from the north-east side
of the central region, and curving toward the south (around -14 -- 16 km s$^{-1}$ in the channel map of Figure \ref{FIG_channel}).
 We call this structure \emph{the south spiral}.

The molecular gas of the south spiral may be spatially distributed because the emission is not prominent in the CARMA-alone map,
while it does appear very clearly in the CARMA+NRO45 map.
The south spiral has lower $R_{\mathrm{10}}$ (Figure \ref{FIG_ratiomap}) than other defined structures.
The southern ridge  appears fragmented, and there are multiple emission peaks in the region.
Figure \ref{FIG_ratiomap} shows  that these peaks have high $R_{\mathrm{10}}$ of  which are comparable to those in the northern ridge.
Therefore, we consider the southern ridge as a counterpart of the northern ridge, even though the southern ridge is more fragmented.

In what follows, we will separate our analyses as \emph{spatial resolved} analyses and  \emph{unresolved} analyses.
For the former, we separate the galactic structures as defined above, while for the latter we average them out with a radial bin and discuss only radial variations.

\section{Physical Properties of Molecular Gas}
\label{sec_physical}
 \begin{table*}
\centering
\caption{$R_\mathrm{10}$ ($I_{\mathrm{^{12}CO(1-0)}}/I_{\mathrm{^{13}CO(1-0)}}$) and $R_\mathrm{21}$ ($I_{\mathrm{^{12}CO(2-1)}}/I_{\mathrm{^{12}CO(1-0)}}$) of each galactic structure and the results of LVG calculations of volume density $n\mathrm{_{H_{2}}}$ and kinetic temperature $T_{\mathrm{k}}$   based on the two observed line ratios.  Note that $R_\mathrm{10}$ is derived based on an angular resolution 3.8$\arcsec$,  $R_\mathrm{21}$ is 20$\arcsec$. $I_{\mathrm{HCN}}$ is the single dish integrated intensity of HCN (1--0) at each position in K km s$^{-1}$.  $L'_{\mathrm{HCN}}$/$L'_{\mathrm{^{12}CO}}$ indicates the relative intensity of dense gas and regular gas tracers. $L'_{\mathrm{HCN}}$/$L'_{\mathrm{^{12}CO}}$ is computed based on single dish observations of HCN and $^{12}$CO (1--0). The single dish beamsize of these observations is about 20$\arcsec$.}
\begin{tabular}{l*{6}{c}r}
\hline
            &   $R_\mathrm{10}$   &   $R_\mathrm{21}$ & ($T_{\mathrm{k}}$, $n\mathrm{_{H_{2}}}$)&  $I_{\mathrm{HCN}}$& $L'_{\mathrm{HCN}}$/$L'_{\mathrm{^{12}CO}}$\\       
\hline
\multirow{2}{*}{central region}        & \multirow{2}{*}{17}& \multirow{2}{*}{$\sim$ 1.0}& ($>$ 40 K, $\sim$ 10$^{3.5}$ cm$^{-3}$) & \multirow{2}{*}{18.9 $\pm$ 1.4 }& \multirow{2}{*}{0.111 } \\
             &  &  & ($\sim$ 20--35 K, $>$ 10$^{4.0-4.5}$ cm$^{-3}$)& &\\
northern ridge             & 12--20 & $\sim$0.8  & ($>$ 15--20 K, $\sim$ 10$^{2-3}$ cm$^{-3}$)   & 4.7 $\pm$ 0.8 & 0.065 \\
southern ridge     &  10--13  &$\sim$0.8 & ($>$ 10 K, $\sim$ 10$^{2-3}$ cm$^{-3}$) & 4.2 $\pm$ 0.8 &   0.051\\
south spiral           & 6--10  &$\leq$0.8 & ($<$ 20 K, $<$ 10$^{3.0}$ cm$^{-3}$) &  $\dots$&  $\dots$ \\
\hline
\label{TAB_lvg}
\end{tabular}
\end{table*}

In order to constrain the spatial resolved temperature and density of molecular gas, we adopt the one-zone Large Velocity Gradient  (LVG) model \citep{Sco74,Gol74}. 
We briefly explain the LVG model (Section \ref{sec_lvg}) and apply it to individual regions (Section \ref{sec_lvg_appli}).
We then compare the derived gas physical properties with a color in infrared.

\subsection{The Large Velocity Gradient (LVG) Model}
\label{sec_lvg}
Molecular emissions depend primarily on three parameters: kinetic temperature ($T_{\mathrm{k}}$) and volume density ($n\mathrm{_{H_{2}}}$) determine the excitation condition, and the optical depth $\tau$ (or alternatively the column density $N_\mathrm{CO}$ per unit velocity width $dv$,  i.e., $N_\mathrm{CO}/dv$)
is important for the radiative transfer. In addition, photon trapping is usually included, with which large $\tau$ effectively reduces the spontaneous
emission rate and affects the excitation condition as well.
The LVG model calculates emission line strengths when ($N_\mathrm{CO}/dv$, $T_{\mathrm{k}}$, $n\mathrm{_{H_{2}}}$) are given.
In reverse, we constrain ($T_{\mathrm{k}}$, $n\mathrm{_{H_{2}}}$) using the observed emission line ratios,
and in our case, $^{12}$CO(1-0), $^{12}$CO(2-1), and $^{13}$CO(1-0).
We employ the LVG code used in \citet{Kod12} with CO-H$_{2}$ collisional cross-sections from \citet{Yan10}. 

For simplicity we fix a possible range of $\log$($N_\mathrm{CO}/dv$)=16.6 -- 17.3 cm$^{-2}$ (km s$^{-1}$)$^{-1}$
as found in the Galaxy and M51 \citep{Sol87,Rod06,Sch10,Kod12}.
NGC 6946 has a similar range  based on the resolved giant molecular clouds analysis \citep{Don12}.
The abundance ratios, [$^{12}$CO/H$_{2}$] and [$^{12}$CO]/[$^{13}$CO], are fixed to the Galactic values
and are 8.0 $\times$ 10$^{-5} $, and 60, respectively \citep{Fre82,Sch10,Kod12}.

\begin{figure*}
    \begin{minipage}{0.45\textwidth}
        \centering
		\includegraphics[width=0.9\textwidth]{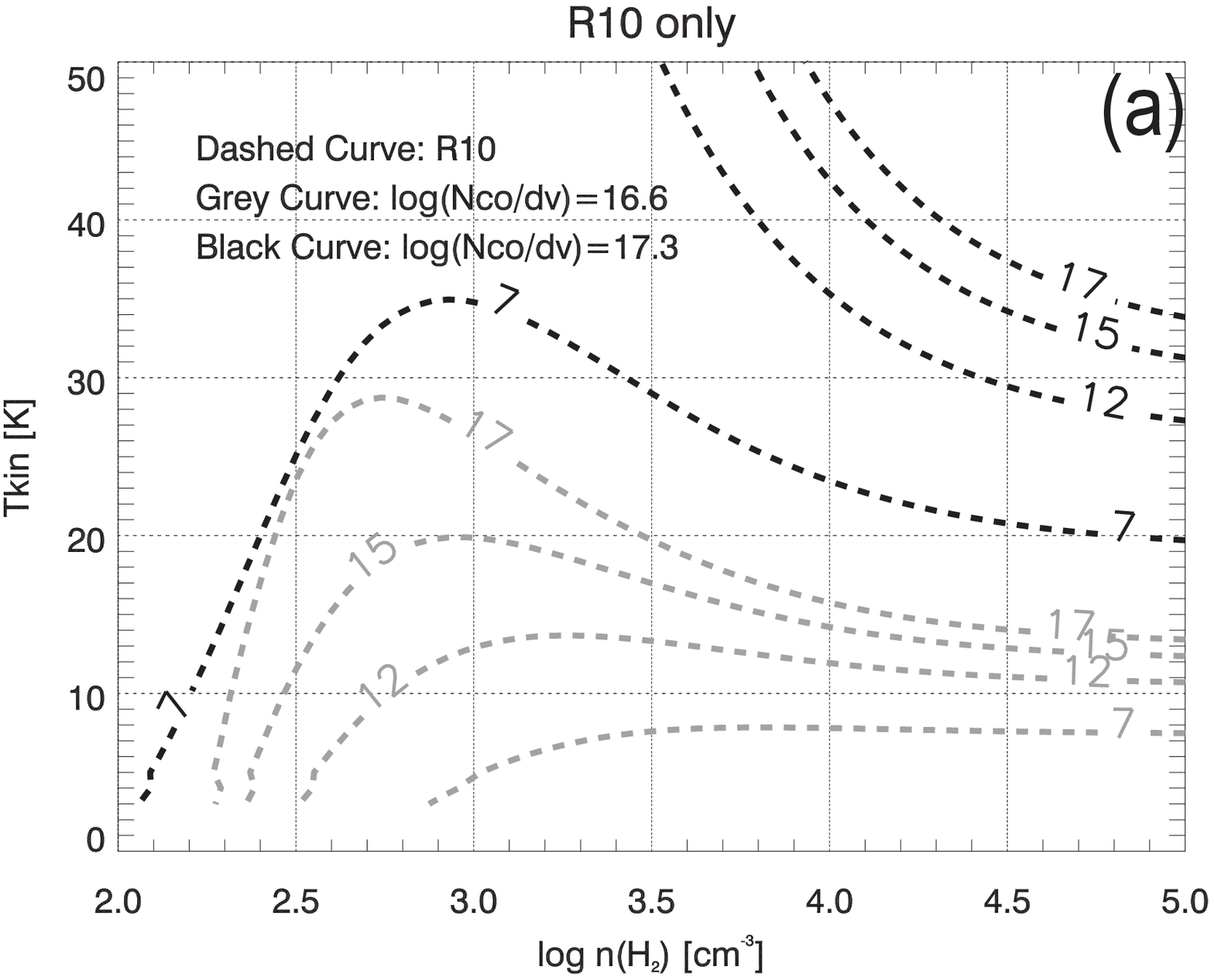}
    \end{minipage}
    \begin{minipage}{0.45\textwidth}
        \centering
		\includegraphics[width=0.9\textwidth]{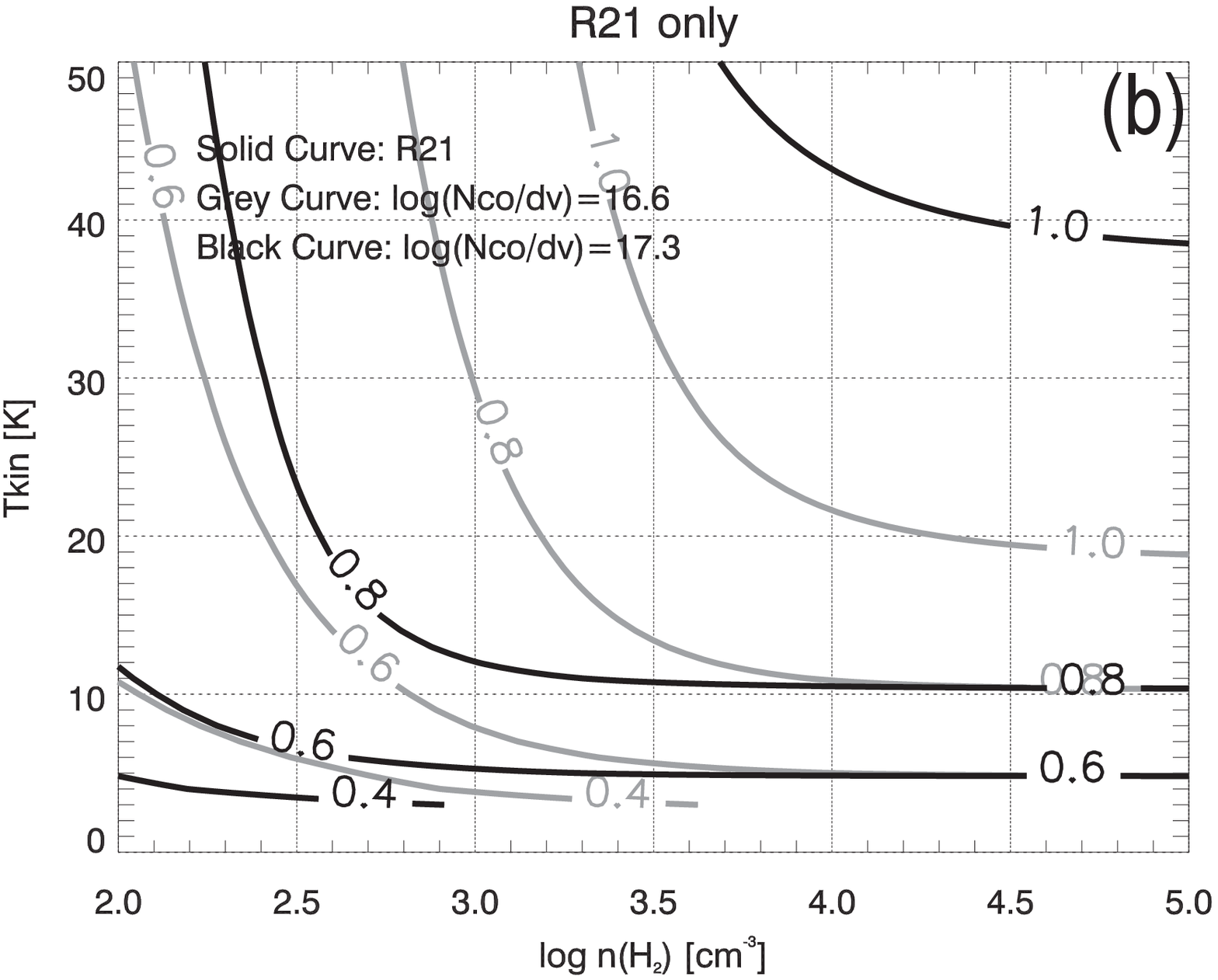}
	    \end{minipage}	
    \caption{Results of Large Velocity Gradient (LVG) calculations. The results are illustrated  as gas temperature ($T\mathrm{_{k}}$) versus density ($n\mathrm{_{H_{2}}}$). Panel (a) and (b) show the results of line ratios of  $R_\mathrm{10}$ and $R_\mathrm{21}$, respectively.   Each ratio have grey and black curves, representing  $\log$($N_\mathrm{CO}/dv$) $=$ 16.6 and 17.3 cm$^{-2}$ (km s$^{-1}$)$^{-1}$, respectively.  }
   \label{FIG_lvg1}
\end{figure*}

Figure \ref{FIG_lvg1} demonstrates the solutions from LVG for (a) $R_\mathrm{10}$ and (b) $R_\mathrm{21}$.
The gray and black lines are for $\log$($N_\mathrm{CO}/dv$) $=$ 16.6 and 17.3 cm$^{-2}$ (km s$^{-1}$)$^{-1}$, respectively.

The LVG results appear non-linear in Figure \ref{FIG_lvg1} due to a mixture of different physical factors. For example, the higher abundance of $^{12}$CO over $^{13}$CO results in a significant difference in optical depth ($\tau_{\rm 12CO} >> \tau_{\rm 13CO}$). Hence, photons from $^{12}$CO are absorbed (trapped) more dominantly in the region where they are emitted, resulting in a lower effective spontaneous emission rate, and thus in a lower critical density ($\sim$300 cm$^{-3}$ over $\sim$2000 cm$^{-3}$. The difference in the critical densities changes their line ratio $R_\mathrm{10}$ dramatically up to $n\sim$2000 cm$^{-3}$. Beyond this density both $^{12}$CO and $^{13}$CO excitations are saturated and optically thick, and $R_\mathrm{10}$ does not depend on the density (Figure \ref{FIG_lvg1}a). An additional complexity is that the optical depth depends on velocity line width, as well as the column density at each $J$ level (which depends on temperature and density). The Doppler broadening of molecular gas is typically wider than the thermal line width, and thus photons from behind may not be absorbed by the gas in front at a different velocity. The LVG models take these into account and show the non-linear lines in Figure \ref{FIG_lvg1}. Observations of $R_\mathrm{10}$ and $R_\mathrm{21}$ enclose an area in the parameter space and give constraints on the physical parameters.

\subsection{Applications to Individual Regions}
\label{sec_lvg_appli}

Figure \ref{FIG_lvg2} shows constraints from the observed $R_\mathrm{10}$ and $R_\mathrm{21}$ in the central region, ridges, and south spiral, and the enclosed areas (gray) indicate the possible ranges of gas temperature and density.
We adopted $\log$($N_\mathrm{CO}/dv$)  $=$ 16.6 to 17.3 as its possible range.

\subsubsection{The Central Region}
\label{sec_lvg_cen}
\begin{figure*}
    \begin{minipage}{0.45\textwidth}
        \centering
		\includegraphics[width=0.9\textwidth]{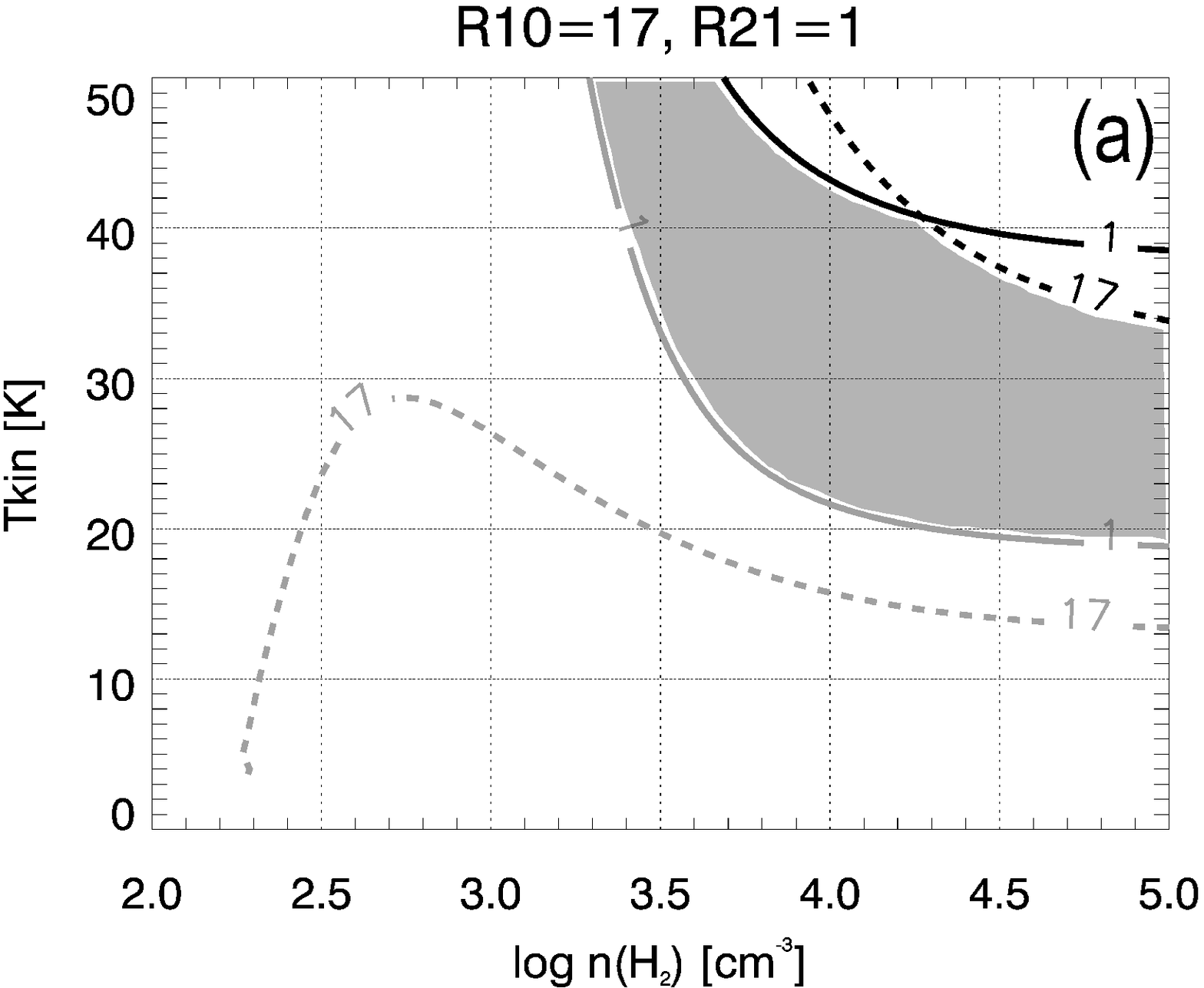}
    \end{minipage}
    \begin{minipage}{0.45\textwidth}
        \centering
        \hspace{-95pt}
		\includegraphics[width=0.9\textwidth]{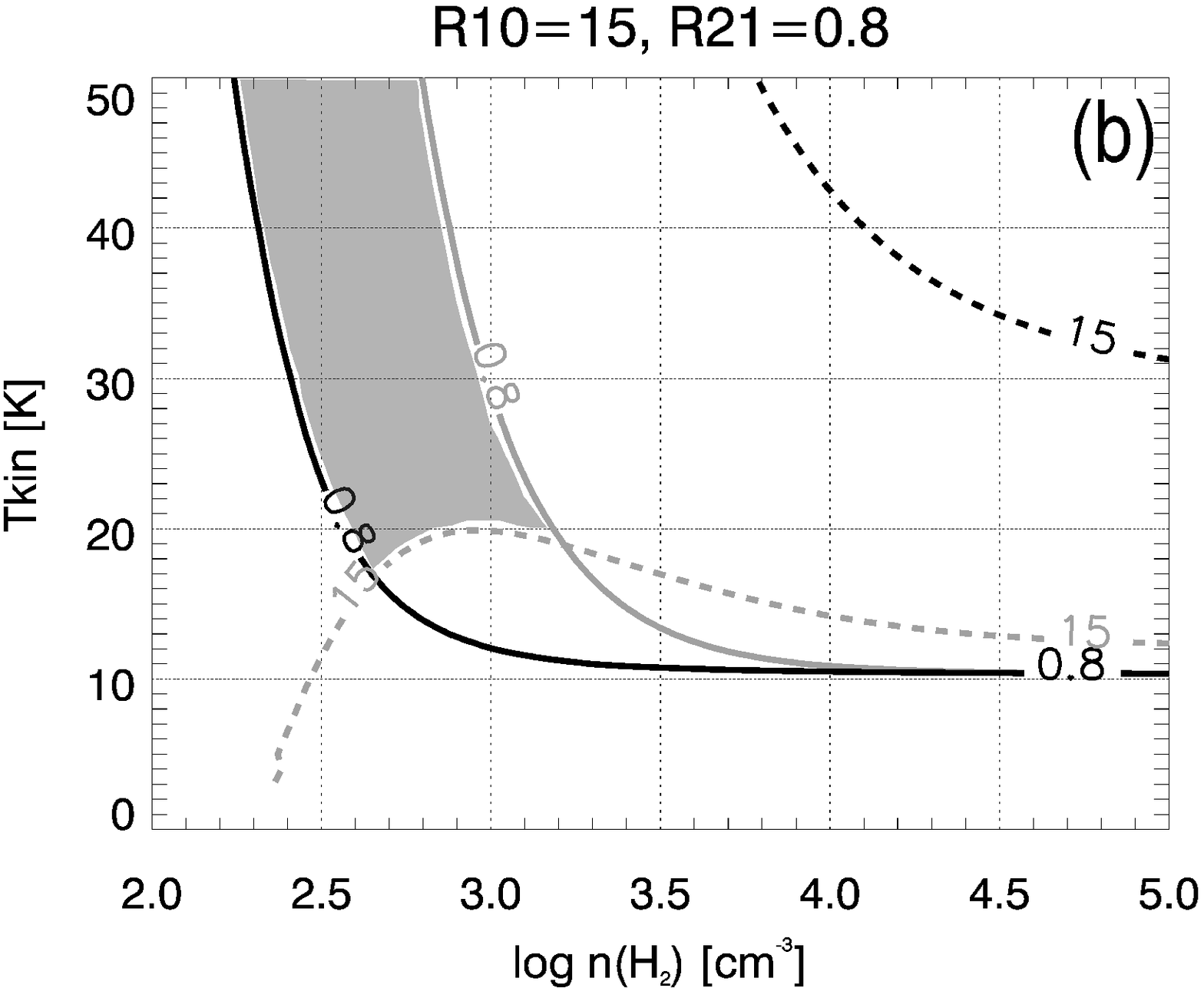}
	    \end{minipage}
	        \begin{minipage}{0.45\textwidth}
        \centering
		\includegraphics[width=0.9\textwidth]{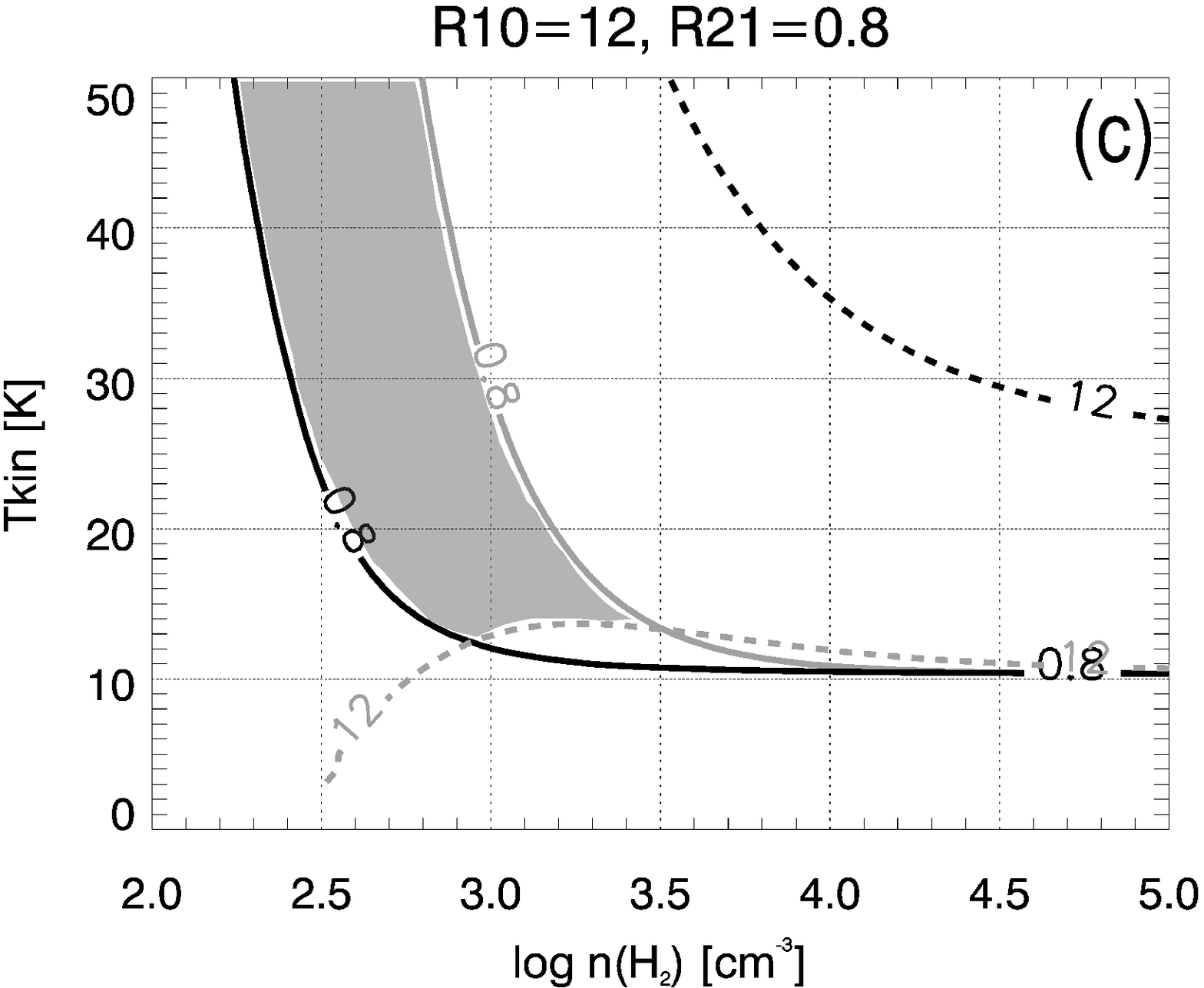}
    \end{minipage}
        \begin{minipage}{0.45\textwidth}
        \centering
		\includegraphics[width=0.9\textwidth]{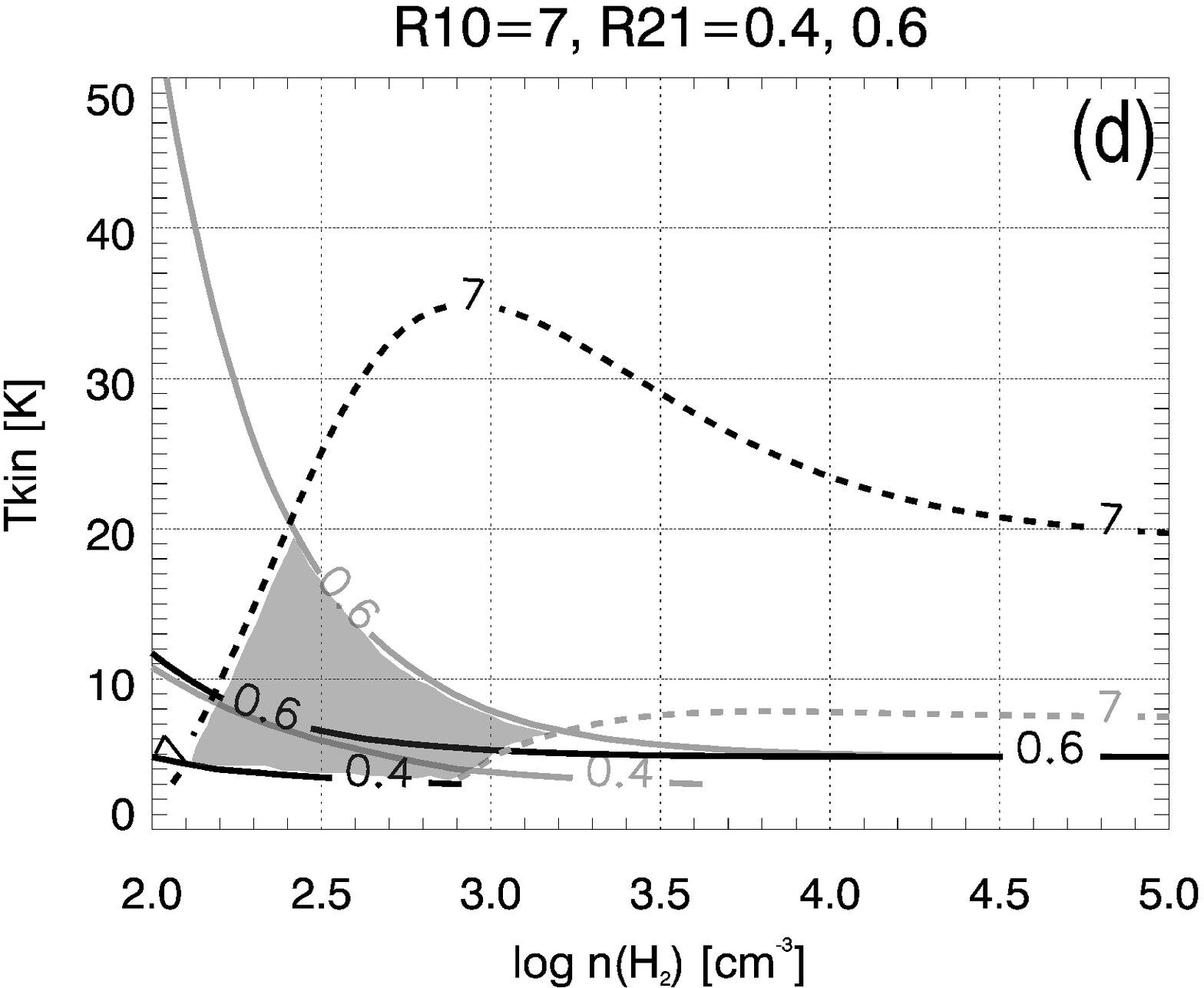}
    \end{minipage}	
    \caption{Results of LVG calculations with combinations of $R_\mathrm{10}$ (dashed curves) and $R_\mathrm{21}$  (solid curves)  for the defined galactic features in \S\ref{sec_name_features}. In each panel, grey and black curves represent $\log$($N_\mathrm{CO}/dv$) $=$ 16.6 and 17.3 cm$^{-2}$ (km s$^{-1}$)$^{-1}$, respectively.  Possible solutions of $T\mathrm{_{k}}$ and $n\mathrm{_{H_{2}}}$ are enclosed within the curves of two line ratios and their two $\log$($N_\mathrm{CO}/dv$), highlighted with grey shadow. (a) Results of the central region where the average $R_\mathrm{10}$ $=$ 17 and the average $R_\mathrm{21}$ $=$ 1.0.    (b) Results of the northern ridge where the average $R_\mathrm{10}$ $=$ 15 and the average $R_\mathrm{21}$ $=$ 0.8. (c)  Results of the southern ridge where the average $R_\mathrm{10}$ $=$ 12 and the average $R_\mathrm{21}$ $=$ 0.8. (d)  Results of the south spiral where the average $R_\mathrm{10}$ $=$ 7 and the average $R_\mathrm{21}$ $\leqslant$ 0.6. }
   \label{FIG_lvg2}
\end{figure*}

At the central region, either temperature or density needs to be high:  ($T_{\mathrm{k}}$ $>$ 40 K, $n\mathrm{_{H_{2}}}$ $\approx$ 10$^{3.5}$ cm$^{-3}$) or ($T_{\mathrm{k}}$ $\approx$ 20 -- 35 K, $n\mathrm{_{H_{2}}}$ $>$ 10$^{4.0-4.5}$  cm$^{-3}$) to satisfy $R_\mathrm{10}\sim$ 17 and $R_\mathrm{21}\sim$ 1.0 (Figure \ref{FIG_lvg2}a).
These solutions can co-exist, and indeed, a presence of a range of gas temperature and density was suggested by \cite{Man13} from their ammonia  (NH$_{3}$) observations (i.e., molecular cloud thermometer). Their reported temperatures, 25 $\pm$ 3 K and 50 $\pm$ 10 K, are consistent with ours.

The derived densities ($\gtrsim 3\times 10^{3}$ cm$^{-3}$) are an order of magnitude higher than the critical density of CO(1-0) excitation, indicating an overall high density in the central region. Although the structures are not resolved with our large beam (hundreds of parsecs), the average high density likely indicates that dense cores and associated star formation exist in this environment.
In fact, the observed $L'_{\mathrm{HCN}}$/$L'_{\mathrm{CO}}$ ratio of the central region is consistent with that of  LIRGs and ULIRGs (\S\ref{sec_hcn_results}), implying that the fraction of dense gas in this region is likely similar to that in starburst galaxies.

\subsubsection{The Ridges and the South Spiral}
\label{sec_lvg_hcncoratio}
Molecular gas at the ridges are  likely cooler and less dense than the galactic center.
The line ratios of   $R_\mathrm{10}$ $\approx$ 15 and $R_\mathrm{21}\approx 0.8$) indicate an average temperature and density of  $>$ 15 -- 20 K and 10$^{2-3}$ cm$^{-3}$, respectively (Figure \ref{FIG_lvg2}b).
The southern ridge  can not be clearly identified in the single dish $R_\mathrm{21}$ map but its gas composition may resemble the northern ridge due to the compatible  $L'_{\mathrm{HCN}}$/$L'_{\mathrm{^{12}CO}}$ and $R_\mathrm{10}$.
Assuming that $R_\mathrm{21}$ of the southern ridge is  $\sim$ 0.8 as the northern ridge, and  its $R_\mathrm{10}$ is about 12, LVG calculations suggest $T_{\mathrm{k}}$ $>$  10 K and $n\mathrm{_{H_{2}}}$ $\approx$ 10$^{2-3}$ cm$^{-3}$ (Figure \ref{FIG_lvg2}c).

South spiral likely has the lowest temperature and density among the regions of interest.
It is reported by previous HCN mapping observations that its $L'_{\mathrm{HCN}}$/$L'_{\mathrm{^{12}CO}}$ is considerably lower than the ridges and the center \citep{Lev08}, indicating that the overall density and the fraction of dense gas is unlikely larger than those areas, i.e., $<$ 10$^{3}$ cm$^{-3}$.
With $R_\mathrm{10}$ $\approx$ 7,  Figure \ref{FIG_lvg2}d therefore suggests that the $R_\mathrm{21}$ is likely less then 0.6.
These ratios give the  solution of temperature of $T_{\mathrm{k}}$ $<$  20 K.

\subsection{Relative Temperature from Infrared Color}
\label{sec_rel_temp}
Dust temperature provides an indirect measure of gas temperature.
Since the LVG calculations only suggest lower or upper limits of gas temperatures due to the non-closed contours, we use an independent methods to constrain the relative temperatures among the region of interests.
Gas temperature ($T_{\mathrm{g}}$) and dust temperature ($T_{\mathrm{d}}$) are not coupled except in very high density regions, but they are positively correlated \citep[e.g.,][]{For14}. We therefore assume that $T_{\mathrm{d}}$ can be used to infer the variation of $T_{\mathrm{g}}$.

\begin{figure*}
    \begin{minipage}{0.45\textwidth}
            \centering
		\includegraphics[width=0.9\textwidth]{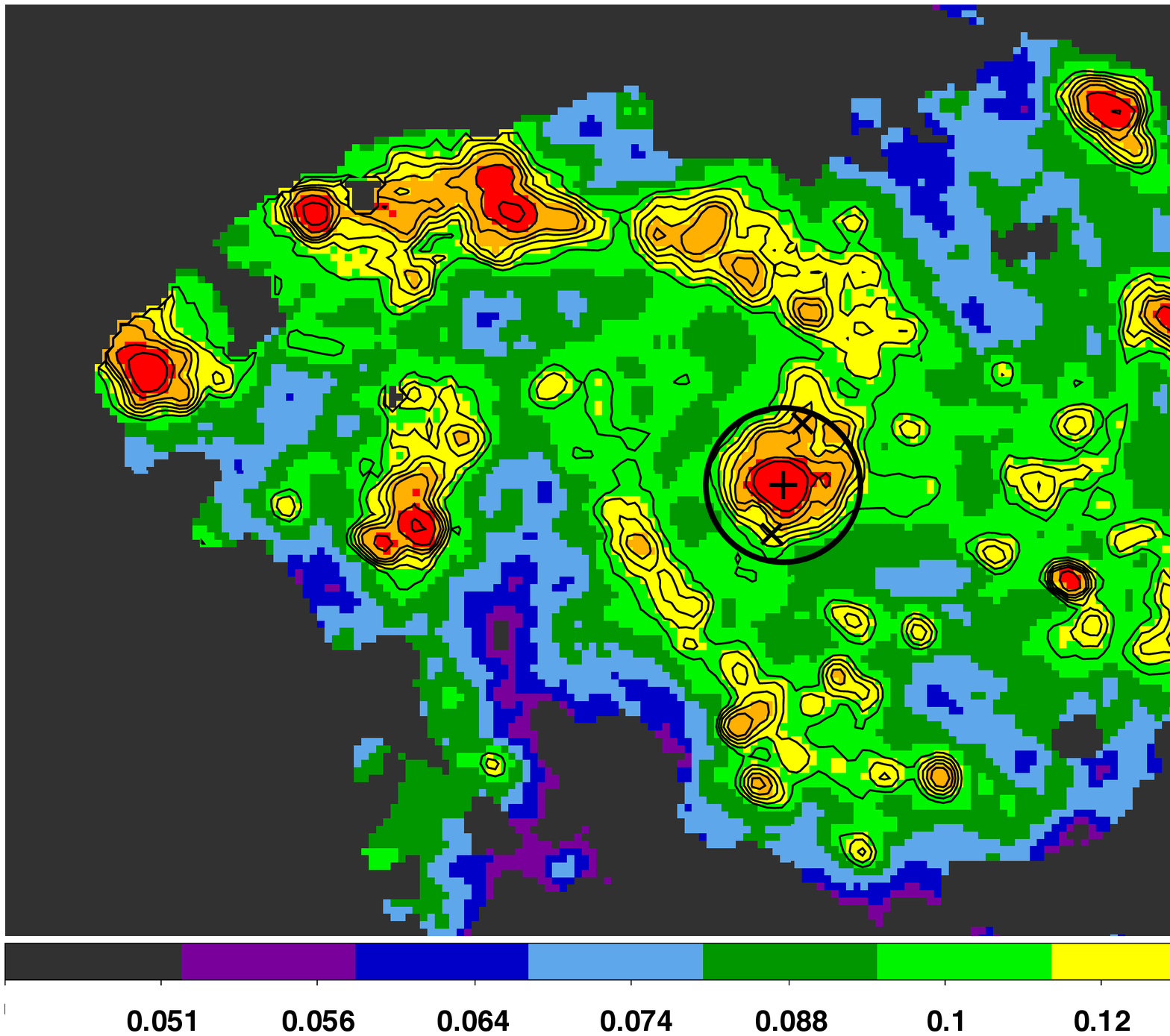}
    \end{minipage}
    \begin{minipage}{0.45\textwidth}
        \centering
                \hspace{-95pt}
		\includegraphics[width=0.9\textwidth]{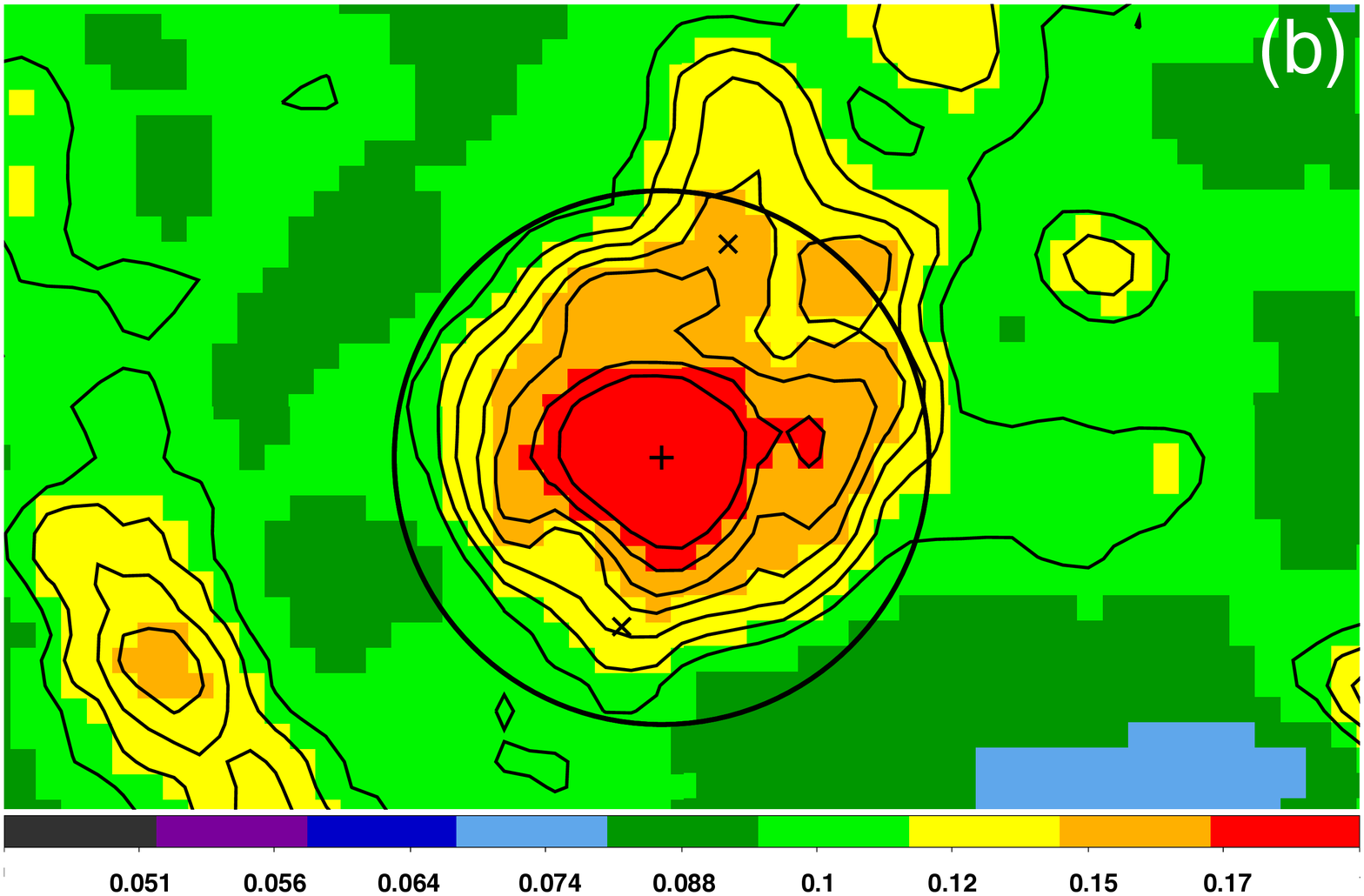}
	    \end{minipage}
	        \begin{minipage}{0.45\textwidth}
        \centering
		\includegraphics[width=0.9\textwidth]{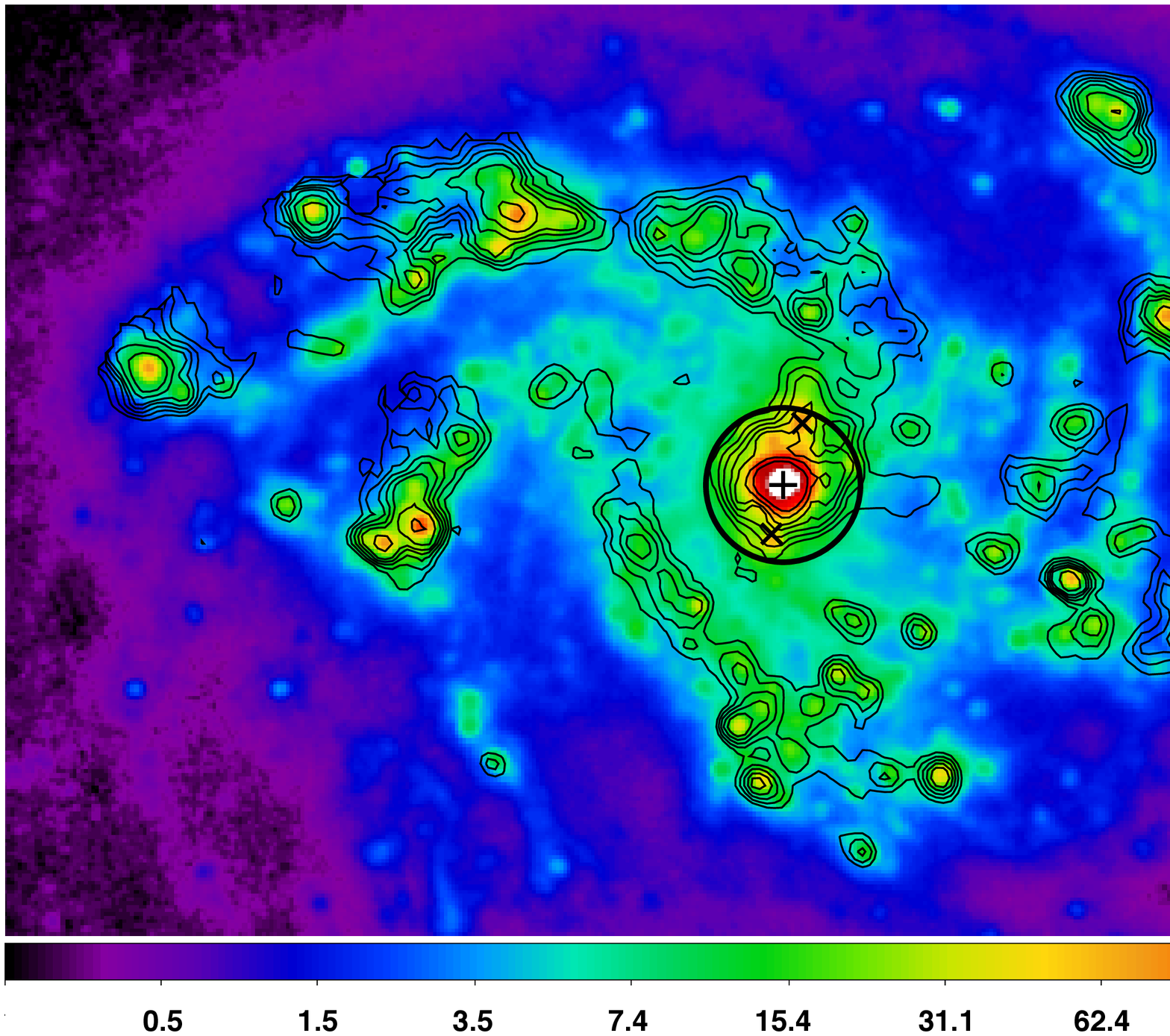}
    \end{minipage}
        \begin{minipage}{0.45\textwidth}
        \centering
		\includegraphics[width=0.9\textwidth]{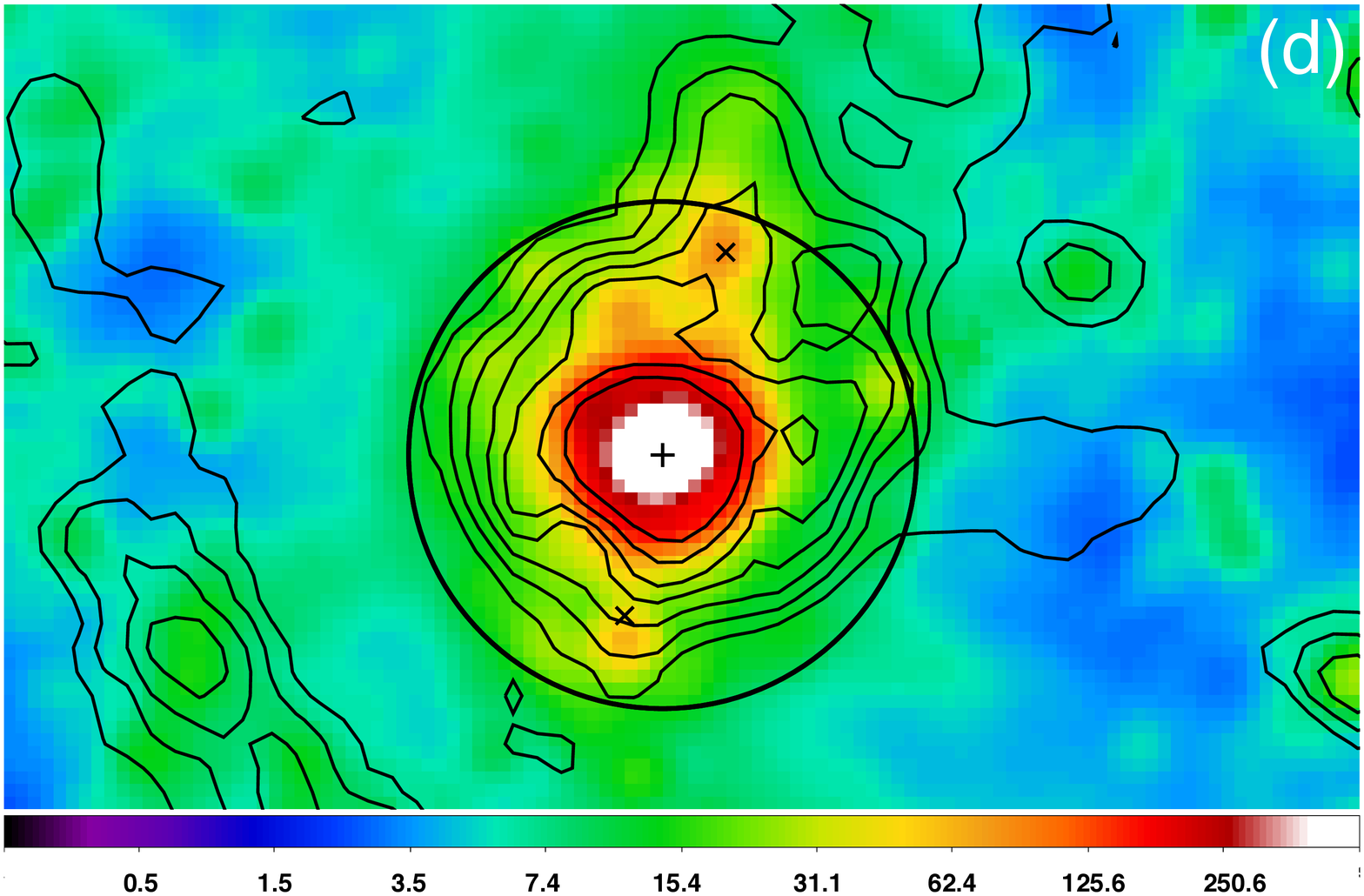}
    \end{minipage}	
    \caption{(a) Infrared color of $S\mathrm{_{70}}/S\mathrm{_{160}}$ (flux ratio of 70$\mu$m to 160$\mu$m) in contours and color scale. Contours are shown at levels of  0.11, 0.12, 0.13, 0.14, 0.15, 0.17, and 0.20.  Black circle indicates the area of  central 1$\arcmin$. Cross marks the galactic center. Locations of the northern and southern ridge are marked with symbol $\times$  (positions of our HCN observations).  PSF of 12$\arcsec$ is indicated at the lower-right corner. (b) A zoom-in view of panel (a). Symbols and color are the same as panel (a). (c) Infrared color (contours) overlaid on star forming regions indicated by Spitzer 24$\mu$m (color scale). Level of the contours are the same as in panel (a).  (d)  A zoom-in view of  panel (c). Symbols and color are the same as panel (c).  }
   \label{FIG_flux_ratio}
\end{figure*}

Infrared color of  70 $\mu$m to 160 $\mu$m flux ratio ($S\mathrm{_{70}}/S\mathrm{_{160}}$) is used to inferred relative $T_{\mathrm{g}}$ among the regions of interest through the SED implied $T_{\mathrm{d}}$, i.e., $S\mathrm{_{70}}/S\mathrm{_{160}}$ increases with $T_{\mathrm{d}}$ (and therefore $T_{\mathrm{g}}$ based on the assumption above).
$S\mathrm{_{70}}/S\mathrm{_{160}}$ map of the entire galaxy is shown in Figure \ref{FIG_flux_ratio}a, while Figure \ref{FIG_flux_ratio}b displays the color map of the central 1$\arcmin$.
The flux ratio is shown with both color scale and contours.
Galactic center is marked with a plus symbol. The black circle denotes the central 1$\arcmin$. 
Two crosses ($\times$) indicate the positions of  HCN observations  at the ridges.

The relative temperatures among the regions of interest  are  consistent with the results of LVG calculations.
$S\mathrm{_{70}}/S\mathrm{_{160}}$  peaks at the galactic center ($\sim$ 0.31), indicating the warmest gas in the galaxy.
The northern ridge ($S\mathrm{_{70}}/S\mathrm{_{160}}$ $\approx$ 0.16) is slightly warmer than the southern ridge and the south spiral ($S\mathrm{_{70}}/S\mathrm{_{160}}$ $\approx$ 0.13).

The temperature distribution of NGC 6946 is likely driven by star formation.
Spatial correlation between $S\mathrm{_{70}}/S\mathrm{_{160}}$ and star forming regions traced by  24 $\mu$m is observed  (Figure \ref{FIG_flux_ratio}c and \ref{FIG_flux_ratio}d).
The  highest $S\mathrm{_{70}}/S\mathrm{_{160}}$ and the strongest 24$\mu$m emission are observed at the central region.
The northern and the southern ridges  characteristic of two stronger 24$\mu$m emission (around the positions of HCN observations) have higher temperature than their surroundings, while the south spiral is lack of 24$\mu$m emission.

\section{Radial  Star Formation Efficiency}
\label{sec_sfr}
We discuss the star formation activity in terms of the radial profile of SFE.
Radial variation of SFE is adopted    since the star formation regions do not coexist with molecular gas but shift  azimuthally  towards the leading sides of the offset ridges (dust lanes), as have been reported in many galaxies \citep[e.g.,][]{Rey98,She00,She02,Asi05}. 
We therefore assume that the star formation regions travel simply azimuthally, and these two postulations can be correlated by azimuthally averaging the properties at the same radius.

Definition of SFE is  number of stars formed per year, it could be formulated as SFR/$M_{\mathrm{H_{2}}}$.
SFR is calculated with the luminosity-SFR relation suggested by \cite{Cal07}:
\begin{equation}
\begin{split}
\mathrm{\frac{SFR}{\mathrm{[M_{\odot}\: yr^{-1}]}}} = & 5.45\times 10^{-42}\times \\ & \left (   \frac{L\mathrm{(H\alpha )_{obs}}}{\left [  \mathrm{erg\; s^{-1}}\right ]}+0.031\frac{L\mathrm{(24)} }{\left [  \mathrm{erg\; s^{-1}}\right ]}\right )\times \cos i,
\end{split}
\label{EQU_SFR}
\end{equation}
where $L\mathrm{(H\alpha )_{obs}}$ and $L\mathrm{(24)}$ are the observed luminosity of H$\alpha$ and 24 $\mu$m, respectively, and $i$ is inclinations of the galaxy. 
Molecular gas mass $M_{\mathrm{H_{2}}}$ is derived with $^{12}$CO (1--0)  because it traces entire volume of molecular gas and requires  less assumptions in the derivation.
The conversion between $^{12}$CO (1--0) flux and $M_{\mathrm{H_{2}}}$ is:
\begin{equation}
\begin{split}
\frac{M_{\mathrm{H_{2}}}}{[\mathrm{M_{\sun}}]}= & 4\times10^{-17}\times \\
& \frac{S\mathrm{_{CO}}}{[\mathrm{Jy\: km\: s^{-1}}]}\frac{D^{2}}{\mathrm{[Mpc]}}\frac{X_{\mathrm{CO}}}{\mathrm{[K\: km\: s^{-1}]}},
\label{EQ_Mgas}
\end{split}
\end{equation} 
where $S\mathrm{_{CO}}$ is flux of $^{12}$CO emission,  $D$ is distance of the galaxy,  $X_{\mathrm{CO}}$ is the CO-to-H$_{2}$ conversion (the only assumption in the calculation).  
Conversion factor of $X_{\mathrm{CO}}$ $=$ 1.2 $\times$ 10$^{20}$ cm$^{-2}$ (K km s$^{-1}$)$^{-1}$ is adopted in this work. 
The value is derived from the virial mass of molecular clouds by \cite{Don12}.
SFR and $M_{\mathrm{H_{2}}}$ are   sampled with a step of 5$\arcsec$ (135 pc) to calculate the radial SFE.

\begin{figure}
  \centering
    \includegraphics[width=0.45\textwidth]{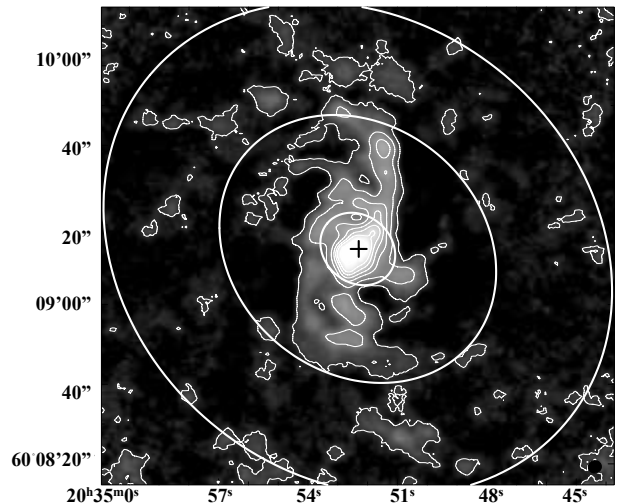}
      \caption{Visualization of  the radial range of galactic environments. $^{12}$CO (1--0) map is shown with greyscale and contours. The galactic center is marked with a  cross. The beamsize is plotted in the lower-right corner. The annuli (or ellipses) take into account the  inclination of 33$^{\circ}$ and position angle of 243$^{\circ}$ of the disk.   The inner  ellipse ($r$ $<$ 10$\arcsec$) denotes the unresolved central region. Offset ridges are enclosed within the middle annulus (10$\arcsec$ $<$ $r$ $<$ 37$\arcsec$),  spiral arms are enclosed within the outer  annulus (37$\arcsec$ $<$ $r$ $<$ 70$\arcsec$). }
      \label{FIG_N6946_ellint}
\end{figure}

We calculate radial SFE in the inner $\sim$ 70$\arcsec$ ($\sim$ 2 kpc) of NGC 6946.
Figure \ref{FIG_N6946_ellint} visualizes the representative  annuli corrected for the inclination and position angle, and shows the  radial extent of galactic structures.
The central unresolved region ($r$ $<$ 10$\arcsec$) are enclosed within the inner ellipse.   This region is discarded in the radial profile of SFE since it is greatly unsolved, but we still calculate the average SFE in this region to be $\sim$ 10$^{9}$ yr$^{-1}$ for comparison. The  middle  annulus (10$\arcsec$ $<$ $r$ $<$ 37$\arcsec$)  corresponds to the offset ridges. Spiral arms emerge  from the offset ridge ends in the outer  annulus of  37$\arcsec$ $<$ $r$ $<$ 70$\arcsec$.

\begin{figure}
  \centering
    \includegraphics[width=0.45\textwidth]{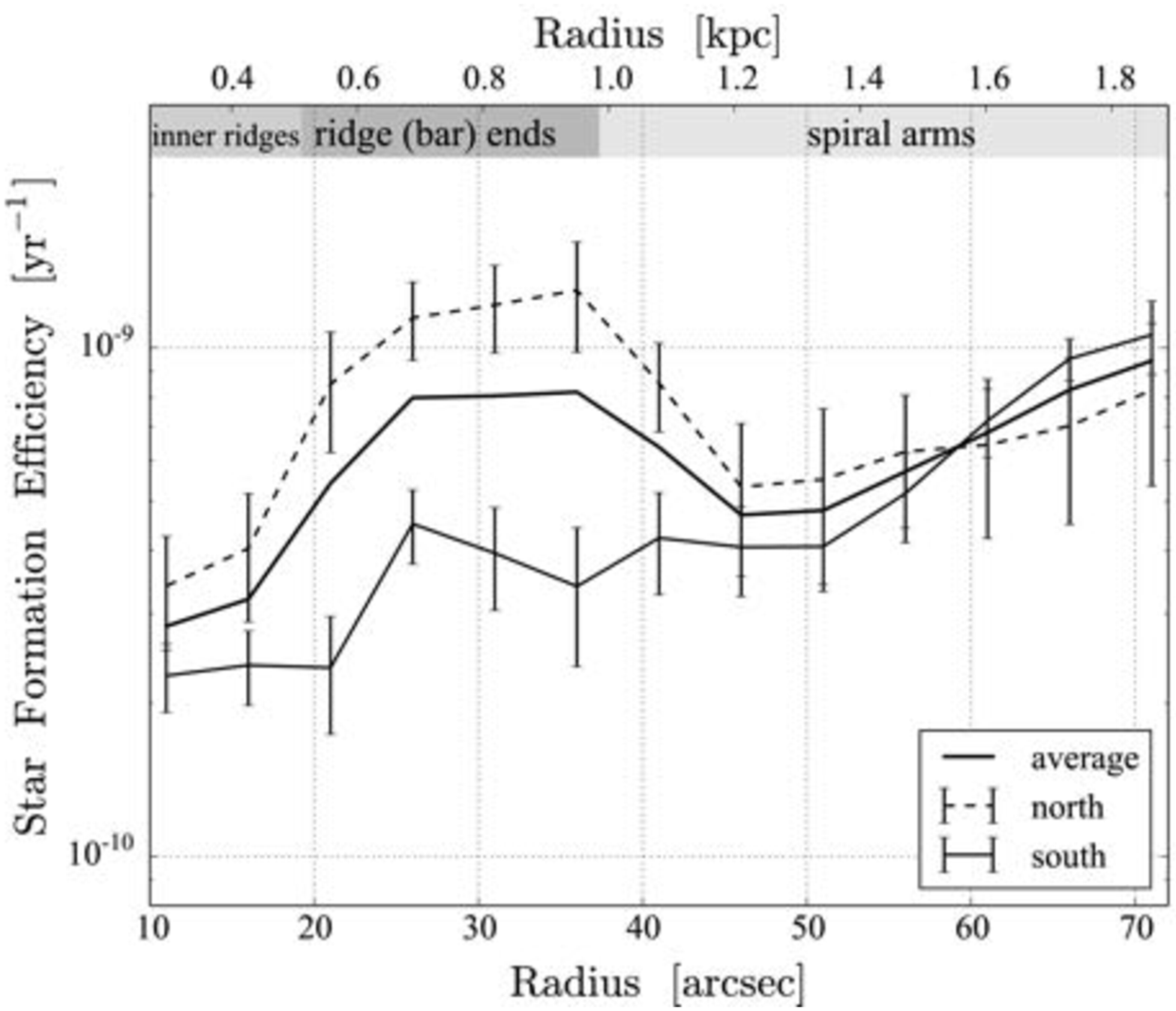}
      \caption{Radial star formation efficiency (SFR/M$_{\mathrm{H_{2},^{12}CO}}$). Dashed, thin-solid and thick-solid curves represent SFE of the northern, southern sides and the average of two sides, respectively.  }
      \label{FIG_SFE_all}
\end{figure}

Figure \ref{FIG_SFE_all} illustrates the radial SFE. 
Dashed, thin-solid and thick-solid curves represent SFE of the northern, southern sides and their average, respectively.
Corresponding galactic features of each radii are indicated at the upper side of the figure.
SFE   varies by $\sim$ 5 times  within the range of $\sim$ 10$^{-10}$ -- 10$^{-9}$ yr$^{-1}$. The range is comparable to  that of nearby disk galaxies seen in pixel-based analysis \citep[e.g.][]{Big08, Ler08,Hua15}.

At the ridges, SFE of the north side is higher than the south side by  $\sim$2 times, but the radial variations of the two sides are similar.
The lowest SFE  ($\sim$ 3 $\times$ 10$^{-10}$  yr$^{-1}$) occurs at the inner ridges, then    increases  to the local maximum at the ridge ends.
The peak SFE is as high as $\sim$ 1.5 $\times$ 10$^{-9}$ yr$^{-1}$ at the northern ridge end, about two times higher than that of the south side.
The variation is consistent to the prediction in terms of the dust lane shocks.  Dust lane shocks increase towards the galactic center along the bar and inhibit  star formation at the inner offset ridges by destroying their parent molecular clouds \citep[e.g.,][]{Rey98,Zur04}.
Additionally, \cite{Wat11} propose that the high SFE at the  ridge ends can be attributed to the increase in the probability of cloud-cloud collisions at this crowded region  that induce star formation.

SFE shows a slight drop beyond the ridge ends but rapidly increases     to $\sim$ 10$^{-9}$ yr$^{-1}$ at $\sim$ 2 kpc, where spiral arms emerge in the both sides.
\cite{Reb12} observed giant molecular cloud associations (GMAs) in the eastern spiral arms of NGC 6946 at $\sim$ 8 kpc away from the galactic center.
They  found a SFE of $\sim$ 10$^{-9}$ yr$^{-1}$ in these spiral GMAs, suggesting that  such high SFE is likely galactic-wide. 
Indeed, optical and infrared observations show that NGC 6946 are filled with massive star-forming regions, super-star cluster, and supernovae  throughout the spiral arms \citep[e.g,][]{Mat97,Lar99,Ced13}.
In spite of that, we note that the analysis of \cite{Reb12} is on the basis of  spatially resolved GMA properties. SFE of   the inner disk certainly requires  high-resolution observations  to further confirm.

Finally, as a caution, we should mention that our analysis of star formation activity has been limited to the radial analysis.
The  derived values of SFE may not be driven by the intrinsic influence of the galactic structures alone because the radial ranges include significant area at the inter-structure regions.  
Significant caution is warranted in interpreting the radial SFE.

\section{Summary}
\label{sec_summary}
In this work, we investigated the physical properties of molecular gas and  star formation activity in the central 2 kpc of NGC 6946. 
 These are done by analyzing the newly observed high-resolution isotopic line $^{13}$CO (1--0) image created by single dish telescope NRO45  and interferometer CARMA,  and other molecular gas tracers in $^{12}$CO (1-0), $^{12}$CO (2-1) and HCN (1-0) (new data from this work), star formation tracers in H$\alpha$ and 24 $\mu$m,  and dust tracers in 60 $\mu$m and 170 $\mu$m from archive (\S\ref{sec_obs_data}). 
Main observational results of the newly observed molecular gas are as follows (\S\ref{sec_results}):
\begin{enumerate}
\item  The NRO45 and NRO45 $+$ CARMA spectra are very similar overall, while the CARMA spectrum has the flux only about 50\% of the NRO45 one.
\item The $^{13}$CO combined  map (our default) shows the central component in details, resolving the central concentration elongated toward north-west to south-east directions, corresponding to the unresolved  nuclear (secondary) bar with major axis of $\sim$ 400 -- 500 pc and the circumnuclear starburst ring.  Emission appears to extend toward north from the nuclear bar reaching the radius of $\sim$ 1 kpc toward north. We call this extension \emph{northern ridge} (offset ridge of molecular gas of the northern primary bar) (\S\ref{sec_name_features}). Emission at the south side is more complex than the north. We define two structures as \emph{southern ridge}  (offset ridge of the southern primary bar) and \emph{south spiral}, which was misunderstood to be part of the southern bar in previous low-resolution images (\S\ref{sec_name_features}). 
\item  Comparison to the archival   $^{12}$CO map (also a combined map of NRO45 and CARMA) shows that the morphology of two CO lines are generally similar, but $^{12}$CO emission  shows more continuous extension overall than  $^{13}$CO.   $^{12}$CO-to-$^{13}$CO ($R_{10}$) ratio varies by a factor of three from the maximum of $\sim$ 17  around the galactic center to the minimum of $\sim$ ${6}$ at the south spiral, covering the large range observed in typical Galactic molecular clouds to starburst galaxies and galaxy mergers. 
\item HCN  single dish observation were made towards  three selected positions: the central region,  northern ridge and the southern ridge to constrain the amount of dense gas. The HCN integrated intensities are 18.9 $\pm$ 1.4, 4.7 $\pm$ 0.8, and 4.2 $\pm$ 0.8 K km s$^{-1}$ at the three regions, respectively.  Comparison between  HCN and CO  shows that their overall spectral profiles are similar to each other, except that the spectra at the southern ridge show a slight difference, where CO observations show multiple peaks while HCN has only one.
\end{enumerate}

Physical properties of  molecular gas are inferred with several methods. The analysis are carried out mainly towards the regions of interest, including the central region, northern ridge, southern ridge, and the south spiral (\S\ref{sec_physical}).
\begin{enumerate}
  \item The one-zone Large Velocity Gradient (LVG) model and the   observed  line ratios of $R_{10}$ and $^{12}$CO (2--1)-to-$^{12}$CO (1--0) ($R_{21}$) are used  to constrain the spatial resolved temperature and density of molecular gas.   LVG calculations show that the bulk molecular gas traced by CO lines in the galactic center is warmer and denser ($\geqslant$ 20 -- 40 K, $n_{\mathrm{H_{2}}}$ $\geqslant$ 10$^{3.5}$ cm$^{-3}$) than that in the offset ridges ($\geqslant$ 10 -- 20 K, $n_{\mathrm{H_{2}}}$ $\approx$ 10$^{2-3}$ cm$^{-3}$). Moreover, the south spiral likely has a  temperature and density lower than the ridges.
  \item  Luminosity ratio of dense gas tracer to low density gas tracer $L'_{\mathrm{HCN}}$/$L'_{\mathrm{^{12}CO}}$ is calculated for the central region and the ridges.  The ratio is often referred to the dense gas fraction. $L'_{\mathrm{HCN}}$/$L'_{\mathrm{^{12}CO}}$ suggests  that   the fraction of dense gas in the central region is  similar to that in starburst galaxies such as LIRGs and ULIRGs, while values of the ridges are close to the global average values of normal galaxies.
  \item   Large-scale temperature distribution is calculated by the infrared color of $S_{70\mu m}$/$S_{160\mu m}$.  Temperature is  spatially correlated with star forming regions seen in 24$\mu$m, suggesting that the large scale temperature distribution of NGC 6946 is  driven by star formation.  Moreover, the relative temperatures among the regions of interest inferred by the infrared color  are  consistent with the results of LVG calculations.  
\end{enumerate}

We discuss the variation of radial star formation efficiency (SFE).
Radial SFE in the inner 2 kpc ($\sim$ 70$\arcsec$) of NGC 6946 is calculated using the gas traced by $^{12}$CO (1--0) (because it can trace the entire volume within the bulk molecular gas) and star formation activity traced by 24$\mu$m and H$\alpha$ (\S\ref{sec_sfr}). 
The key results are as follows.
\begin{enumerate}
\item   SFE of the north side of the galaxy is higher than that of the south side by $\sim$ 2 times. For each side of the galaxy, radial SFE changes by about five times in the inner 2 kpc disk.  
\item  In spite of the different SFE, radial SFE  share similar trend of variation at  the two sides.   Low SFE  is seen  in the radial range of the inner (midway) offset ridges, whereas,  the ridge ends show high SFE.   The  variations of SFE agree with the prediction based on the effect of the dust-lane shocks and the increase in the probability of cloud-cloud collisions in  high-interaction environment.  
\end{enumerate}

\vspace{15pt}
 We thank referee Alessandro Romeo for providing constructive comments which have helped to improve the paper. This work is supported by the Associate Support System of the National Astronomical Observatory of Japan (NAOJ).
JK acknowledges support from the NSF through grant AST-1211680 for the work presented in this paper, as well as NASA through grant NNX09AF40G \&  NNX14AF74G, a Herschel Space Observatory grant, and an Hubble Space Telescope grant (No. 12490).
This research has made use of the NASA/IPAC Extragalactic Database (NED) which is operated by the Jet Propulsion Laboratory, California Institute of Technology, under contract with the National Aeronautics and Space Administration.

\end{document}